\documentclass{aa}
\usepackage{epsfig}
\usepackage{graphics} 
\usepackage{graphicx} 
\usepackage{color, colortbl}
\usepackage{tikz}
\usepackage{pgfplots}
\usepackage[colorlinks=true,     linkcolor=blue, citecolor=blue, filecolor=blue, urlcolor=blue]{hyperref}

\pgfplotsset{compat=1.18}

\begin{document}

\def\kms{\rm km\,s^{-1}}
\def\percc{\rm cm^{-3}}
\def\persqcm{\rm cm^{-2}}
\def\arcsec{^{\prime\prime}}
\def\mum{\mu{\rm m}}

\def\htwo{\rm H_2}
\def\water{\rm H_2O}
\def\formyl{\rm HCO^+}
\def\diaz{\rm N_2H^+}

\thicklines

\title{Oxygen in the protostellar clump OMC-2 FIR4\thanks{Based on observations made with the NASA/DLR Stratospheric Observatory for Infrared Astronomy (SOFIA). SOFIA is jointly operated by the Universities Space Research Association, Inc. (USRA), under NASA contract NNA17BF53C , and the Deutsches SOFIA Institut (DSI) under DLR contract 50 OK 2002 to the University of Stuttgart. The publication uses data collected with the Atacama Pathfinder Experiment (APEX, programme ID 0105.C-0281A-2020). The APEX Project is led by the Max Planck Institute for Radio Astronomy at the ESO La Silla Paranal Observatory.}}

  \author{J. Harju
          \inst{1,2}
          \and
                    P. Caselli \inst{1}
            \and
                   R. G{\"u}sten \inst{3}  
                 \and
                    H. Wiesemeyer \inst{3}
                    \and
                   S. Br{\"u}nken \inst{4,5} 
            \and
                    M. Mertens \inst{6}                 
            \and          
                    O. Sipil{\"a} \inst{1}        
          \and 
                 J. Stutzki \inst{6}
                             \and
                    F. Wyrowski \inst{3}    
}
          
\institute{Max-Planck-Institut f{\"u}r extraterrestrische Physik, Gie{\ss}enbachstra{\ss}e 1, D-85748 Garching, Germany
\and
Department of Physics, P.O. Box 64, FIN-00014, University of Helsinki, Finland
\and
    Max-Planck-Institut f{\"u}r Radioastronomie,
    Auf dem H{\"u}gel 69, D-53121 Bonn, Germany
\and
HFML-FELIX, Toernooiveld 7, 6525 ED Nijmegen, The Netherlands
\and
Institute for Molecules and Materials, Radboud University, Heyendaalseweg 135, 6525 AJ Nijmegen, The Netherlands
\and
I. Physikalisches Institut, Universit{\"a}t zu K{\"o}ln, Z{\"u}lpicher Str. 77, 50937 K{\"o}ln, Germany}

\date{}
\abstract{Atomic oxygen (\ion{O}{i}), OH, $\water$, and CO are the main carriers of oxygen in dense interstellar gas. The far-infrared lines of these species are considered important coolants of shock-heated dense gas associated with protostellar outflows. Their relative abundances depend on the nature of the shock and usually remain uncertain in observational studies.}
{We determine the relative abundances of \ion{O}{i}, OH, $\water$, and CO in the warm inner parts of the protostellar clump OMC-2 FIR4 in Orion A. The clump contains several young stellar objects.}
{The upGREAT receiver including the High Frequency Array (HFA, operating at 4.74\, THz, $\lambda=63\,\mum$) onboard the Stratospheric Observatory for Far-Infrared Astronomy (SOFIA) was used to observe OMC-2 FIR4 in the lines of \ion{O}{i}, OH, OD, HDO, and CO. Additional HDO lines were observed with the Atacama Pathfinder Experiment (APEX). Archival $\water$ and CO spectra observed by the {\sl Herschel} satellite were included in the analysis. The observed lines were reasonably well reproduced by an expanding spherical shell model.}
{The \ion{O}{i} spectrum at $63\,\mum$ towards OMC-2 FIR4 is dominated by a broad line component, on top of which medium-wide and narrow line components can be discerned. The same components are present in the OH, $\water$, and high-$J$ CO spectra towards this source. We find that \ion{O}{i} is more abundant than $\water$ in the shocked gas. In the broad line component, the following abundance ratios are derived: \ion{O}{i}/$\water \sim700$, \ion{O}{i}/OH $\sim300$, \ion{O}{i}/CO $\sim4$. {The high relative abundance of atomic oxygen there} suggests an origin in dissociative J-shocks that are associated with strong ultraviolet radiation. {The \ion{O}{i}/CO ratio decreases below unity in the components with a smaller velocity dispersion, and these components also have higher abundances of $\water$ than the broad line component, although remaining below that of CO.} The HDO/$\water$ ratio in the low-velocity components corresponds to the average ratio in the icy mantles of dust grains, and the presence of water there could also be understood in terms of  sublimation without invoking high-temperature chemistry.}  
{}


\maketitle
\nolinenumbers

\section{Introduction}

Winds and jets from embedded protostars give rise to collisionally heated and accelerated dense gas that can be traced by broad far-infrared (FIR) lines of $\water$, CO, OH and atomic oxygen, \ion{O}{i} (\citealt{2010A&A...521L..30K}; \citealt{2014A&A...562A..45K}; \citealt{2014A&A...572A..21M}; {\citealt{2017A&A...602A...8G}; \citealt{2017A&A...605A..93K}; \citealt{2021A&A...648A..24V} and references therein). These lines are considered major coolants of shocked gas and, therefore, are important for the dynamical evolution of protostellar systems. The relative abundances of the mentioned species and their contributions to the cooling of FIR lines change from source to source \citep{2018ApJS..235...30K}. These changes depend on the nature of the shocks associated with the outflows and, ultimately, on the luminosity and evolutionary stage of the central source and the physical conditions in the envelope. The two main types of shocks, dissociative J-shocks and smoother magnetohydrodynamic C-shocks \citep{1980ApJ...241.1021D}, are predicted to leave their traces in the chemical composition of the gas during and immediately after the shock (e.g. \citealt{1989ApJ...340..869N}, \citealt{1989ApJ...342..306H}; \citealt{1996ApJ...456..250K}; \citealt{2019A&A...622A.100G}; \citealt{2020A&A...643A.101L}).  On the observational side, the abundance of water is typically found to be lower than predicted by chemical shock models according to which all volatile oxygen not locked up in CO goes to $\water$. There are also indications that the cooling by \ion{O}{i} is more important for Class I protostars compared to the earlier Class 0 stage \citep{2018ApJS..235...30K}. Both observational results have been suggested to be traceable to ultraviolet (UV) radiation from the central star(s) or strong shocks impinging on non-dissociative C-shocks (\citealt{2013A&A...552A.141K}; \citealt{2013A&A...557A..23K}; \citealt{2014A&A...572A...9K}; \citealt{2017A&A...605A..93K}). On the other hand, the \ion{O}{i} line at $63\,\mum$ is predicted to be the main coolant of dense dissociative J-shocks \citep{1989ApJ...342..306H}, and contributions from multiple shocks with different strengths are typically needed to explain the observed FIR line ratios. An obstacle to determining the oxygen budget in protostellar outflows is the scarcity of spectrally resolved \ion{O}{i} observations. 
 
Observationally, the analysis is complicated as bright \ion{O}{i} emission also originates in photodissociation regions (PDRs) on the surfaces of dense clouds exposed to strong UV radiation from nearby hot stars. In regions forming stellar clusters it is possible to see \ion{O}{i} fine structure lines arising from two separate components of warm gas: out-flowing gas near the protostars that is heated both mechanically and by radiation, and gas at cloud surfaces heated primarily by the photoelectric effect (\citealt{1997ApJ...481..343H}; \citealt{2016A&A...596A..26G}). An additional difficulty is that \ion{O}{i} lines from massive-star-forming regions are often self-absorbed or absorbed by unrelated foreground clouds (\citealt{2015A&A...584A..70L}; \citealt{2021ApJ...916....6G}; \citealt{2024A&A...690A.294G}).

Here, we present the \ion{O}{i}, OH, and HDO spectra observed with the upGREAT receiver on the Stratospheric Observatory for Far-Infrared Astronomy (SOFIA) towards the strong FIR source OMC-2 FIR4 (\citealt{1990A&A...228...95M}; \citealt{1997ApJ...474L.135C}; \citealt{1999ApJ...510L..49J}) in Orion Molecular Cloud 2 (OMC-2). OMC-2 is part of the integral-shaped filament in the Orion A molecular cloud complex. The distance of the cloud is approximately 390 pc \citep{2017ApJ...834..142K}. In the analysis, we include several $\water$ and CO lines previously observed {in the course of the {\sl Herschel} key programme "Chemical HErschel Surveys of Star-forming regions" (CHESS; \citealt{2010A&A...521L..22C})},  and two HDO lines observed with the Atacama Pathfinder Experiment (APEX). The dense molecular clump OMC-2 FIR4 with a mass of approximately $30\,M_\odot$ dominates the FIR continuum emission longward of $\lambda=160\,\mum$ and is the strongest source of high-excitation FIR lines of $\water$, OH, CO, and \ion{O}{i} in OMC-2 (\citealt{2013ApJ...763...83M}; \citealt{2013A&A...556A..57K}; \citealt{2014ApJ...786...26F}; \citealt{2016A&A...596A..26G}). OMC-2 FIR4 contains several compact millimetre continuum sources (\citealt{2017A&A...600A.141K}; \citealt{2008ApJ...683..255S}; \citealt{2023ApJ...944...92S}) of which four are associated with known young stellar objects (YSOs). These are HOPS-108 and HOPS-64, detected in the course of the {\sl Herschel} Orion Protostar Survey (HOPS; \citealt{2013AN....334...53F}), and the compact radio continuum sources VLA 15 and VLA 16 detected by {\cite{2017ApJ...840...36O}}. The nature of these objects has also been discussed in \cite{2012ApJ...749L..24A}, \cite{2016ApJS..224....5F}, and \cite{2019ApJ...886....6T}. 

The OMC-2 FIR4 clump is exposed to strong FUV radiation from the Orion Nebula Cluster (\citealt{2006A&A...449..609J}; \citealt{2018ApJ...859..136F}), and it is also claimed to be pervaded by an extremely high flux of cosmic-ray-like particles, possibly originating from an internal source (\citealt{2014ApJ...790L...1C}; \citealt{2017A&A...605A..57F}; \citealt{2018ApJ...859..136F}). In addition, several authors have presented evidence that a powerful outflow from the nearby intermediate-mass protostar HOPS-370 located in the OMC-2 FIR3 clump collides with the FIR4 clump, causing shocks that cannot be easily attributed to outflows from protostars inside the clump (\citealt{2008ApJ...683..255S}; \citealt{2016A&A...596A..26G}; \citealt{2022A&A...667A...6C}; \citealt{2023ApJ...944...92S}). According to \cite{2023ApJ...944...92S}, the four embedded YSOs and two other $\lambda=1.3$\,mm sources are, however, associated with compact outflows detected in the CO $J=2-1$ and SiO $J=5-4$ lines (see Figure 10 of \citealt{2023ApJ...944...92S}). In addition, the Class 0 protostar HOPS-108 possibly drives a collimated molecular jet detected in SiO \citep{2023A&A...671A..35L}.

The purpose of the present work is to derive the relative abundances of \ion{O}{i}, OH, $\water$, and CO in the warm inner parts of the OMC-2 FIR4 clump and in the PDR at its surface. The abundance ratios have an immediate connection to the oxygen budget and the role of the FUV field in the two regimes. We employ an expanding spherical shell model for the clump. The expansion, implied by the observed line shapes, is assumed to be driven by winds from the embedded protostars. Although complex and disordered structures are seen in interferometric images of this object (\citealt{2022A&A...667A...6C}; \citealt{2023ApJ...944...92S}), the simple model can produce spectral lines that resemble the single-dish spectra presented here.      

\section{Observations}

\subsection{SOFIA}

The observations towards OMC-2 FIR4 were made during SOFIA Cycle 6 on December 11, 2018 on a flight starting from Palmdale, California. These were part of project 06\_0179  (PI P. Caselli) employing the German REceiver for Astronomy at Terahertz Frequencies (GREAT; \citealt{2012A&A...542L...1H}). The front-end unit consisted of the upGREAT High Frequency Array (HFA; \citealt{2016A&A...595A..34R}) and the 4GREAT (4G) receiver \citep{2021ITTST..11..194D} with four co-aligned pixels at four different frequencies. The HFA was tuned to the frequency of the ground-state fine structure line of \ion{O}{i}, [\ion{O}{i}] ${^3P}_1-{^3P}_2$ at $\sim4.74$\,THz ($63.2\,\mum$). The line was observed in the lower sideband (LSB) of the receiver. The array has seven pixels in a hexagonal geometry. The mean separation between pixels is $13\farcs8$. The footprint of the HFA array superposed on the $100\,\mum$ surface brightness map from {\sl Herschel}/PACS is shown in Fig.~\ref{hfa_on_pacs}. The 4G channels, coaligned with and operating in parallel to the HFA central pixel, were tuned to cover the fundamental rotational lines of OH, OD and HDO at 2.51, 1.39, and 0.89\,THz, respectively, and the $J=5-4$ line of CO at 576 GHz. The OH line at 2.51\,THz was observed in the upper sideband (USB) of the 4G-4 receiver whereas the OD, HDO, and CO lines were placed in the LSB. The observed transitions, their frequencies, and the upper state energies are listed in Table~\ref{obs_lines}. The table also gives the telescope beam sizes at the frequencies used. The pointed observations were targeted at R.A. 05:35:27.0, Dec. -05:09:57 (J2000.0). This position corresponds to the millimetre and submillimetre peak of OMC-2 but lies $3\arcsec$ north of the mid-infrared peak (see Table~4 in \citealt{2014ApJ...786...26F}).

The observations were made in the dual beam switching mode, where the secondary mirror was chopped between the source and two reference positions lying symmetrically to the equatorial east and west from the target. The switching amplitude was $125\arcsec$ (chop throw $250\arcsec$) and the switching frequency was 2.5\,Hz. An on-off cycle was executed by first chopping for 30\,s of time with the reference lying $2\times125\arcsec$ towards positive R.A. (phase A), then moving the telescope to chop phase B for another 30\,s (throw $2\times125\arcsec$ towards negative R.A.). After four repetitions ($4\times2\times30$\,s), a hot + cold load calibration measurement was performed. The total observing time was 125 minutes.

Despite the relatively large chop throw, post-flight data inspection revealed \ion{O}{i} emission in both off-source chop phase signals; this manifested itself in pseudo-self-absorption in the on-source spectra after subtraction of the off-phase signal. However, thanks to the high quality of the deep pointed integration towards OMC-2 FIR4 we were able to recover the underlying spectra. By extracting data of the on- and off-source integrations for each pixel, for each chop phase separately, plus the calibration data on the hot and cold loads, it was possible to reference the off-source signal for each chop phase against the hot load (mimicking "load switching"). This provided the astronomical off-source "contamination" superimposed on the emission spectrum of the FIR sky. The good system stability of the GREAT receivers was beneficial to the process.

Re-processing the data revealed about 2\,K\footnote{The line intensity in this paper is expressed as the main beam brightness temperature, $T_{\rm MB}$, which is the   brightness temperature, $T_{\rm B}$, of the radiation source on the Rayleigh-Jeans scale detected by the telescope, averaged over the solid angle of the beam \citep{2001A&A...365..285B}.} \ion{O}{i} signals in the off-chop positions ($3-4\,\kms$ in width), with slightly different profiles between the two off-positions ($500\arcsec$ apart). We fitted Gaussian profiles to the off-spectra. Here, we co-added the data of all array pixels assuming that the emission does not change significantly over the $27\arcsec$ field-of-view of the HFA. The averaged \ion{O}{i} spectra of the two off-positions are shown in Fig.~\ref{off_spectra} and their line parameters are listed in Table~\ref{off_gaussian}. For five (out of seven) pixels, no difference within the noise could be discerned if this step were performed pixel-by-pixel; the other two were noise limited. The Gaussian profiles were then added (separately for the two chop phases) to the on-source spectra. The corrected \ion{O}{i} spectra towards OMC-2 FIR4 are shown in Fig.~\ref{o_spectra}.  

We inspected the data from the 4G receivers operated in parallel with the HFA and found that the CO $J=5-4$ spectrum was also affected by contamination from the off-position, whereas the other transitions appeared to be clean. We processed the CO data in a similar manner as described above.  The corrected CO $J=5-4$ spectrum is shown in Fig.~\ref{co_spectra} in the Appendix together with three high-$J$ spectra observed with {\sl Herschel}/HIFI. The OD line at 1391.5\,GHz was not detected in the spectrum with an RMS noise level of  0.25\,K ($T_{\rm MB}$). This line is not discussed below.

\begin{table}[ht]
\caption[]{Observed transitions.}
\label{obs_lines}
\begin{tabular}{llrrr} \hline
front-end & transition & frequency & $E_{\rm u}$ & beam \\ 
           &  & (GHz) & (K) & ($\arcsec$)$^\dagger$ \\ \hline
\noalign{\smallskip}
\multicolumn{5}{c}{SOFIA} \\ \hline
\noalign{\smallskip}
HFA$^*$ & [\ion{O}{i}] ${^3P}_1-{^3P}_2$ &  4744.777 & 227.7 & 6.3 \\
4G-4$^\ddagger$& OH $5/2-3/2$ & 2514.316 & 120.7 & 10.6 \\ 
4G-3& OD $5/2-3/2$ &  1391.495 & 66.8 & 20.0 \\ 
4G-2& HDO $1_{11}-0_{00}$ &  893.639 & 42.9 & 28.7 \\
4G-1& CO $5-4$ &  576.268 & 83.0 & 55.2 \\ \hline
\noalign{\smallskip}
\multicolumn{5}{c}{APEX} \\ \hline
\noalign{\smallskip}
nFLASH & HDO $1_{01}-0_{00}$ & 464.925 & 22.3 & 13.4 \\
nFLASH & HDO $2_{11}-2_{12}$ & 241.562 & 95.2 & 25.8 \\ \hline
\noalign{\smallskip}
\end{tabular}

$\dagger$ FWHM 

$^*$ upGREAT High Frequency Array

$\ddagger$ 4GREAT four-detector receiver

\end{table}

\begin{figure}[ht]
\unitlength=1.0mm
\begin{picture}(80,60)(0,0)
\put(0,0){
\begin{picture}(0,0)
\includegraphics[width=8cm,angle=0]{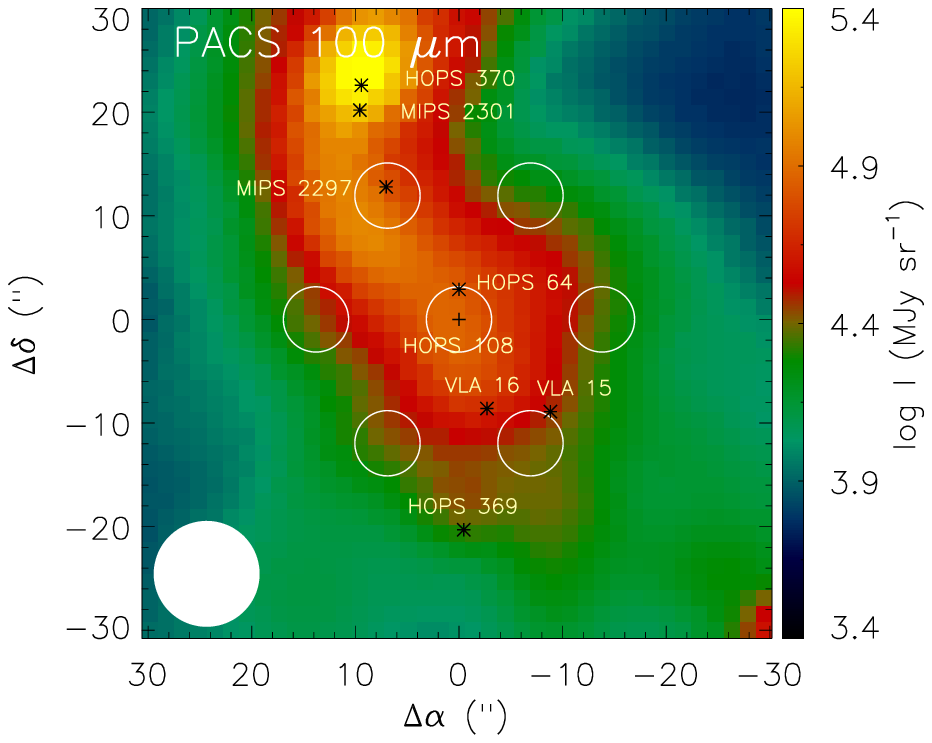}
\end{picture}}
\end{picture}
\caption{Footprint of the upGREAT HFA array on the $100\,\mum$ surface brightness map of OMC-2 FIR4 observed by {\sl Herschel}/PACS. The open circles show the positions of the seven pixels of the HFA and their size corresponds to the FWHM of the beams. The solid white circle shows the approximate beam size of the PACS instrument. The strong source in the north is the protostellar clump OMC-2 FIR3. {The positions of the YSOs associated with FIR3 and FIR4 are indicated. Their coordinates and classifications can be  found in, for example, \cite{2017ApJ...840...36O} and \cite{2019ApJ...886....6T}.}}
\label{hfa_on_pacs}
\end{figure}

\subsection{APEX}

The APEX observations towards OMC-2 FIR4 were made in the course of the programme 0105.C-0281A-2020 (PI P. Caselli). These were single-point position switched observations with the reference chosen $120\arcsec$ east of the on-position given above. The ground-state line of HDO at 465 GHz was observed in the LSB of the nFLASH460 receiver. These observations were obtained in October 2020. The HDO $2_{11}-2_{12}$ line at 242 GHz was observed in April 2021 using the nFLASH230 receiver. The line was placed in the USB. The nFLASH receivers are dual polarisation, dual sideband (2SB) systems, meaning that both the upper and lower sidebands, separated by 12 GHz,  are present in the output in two orthogonal polarizations. The widths of the sidebands are 8\,GHz for nFLASH230 and 4\,GHz for nFLASH460. The bands are recorded with several Fast Fourier Transform spectrometers at a spectral resolution of 61\,kHz. The APEX spectra contain a large number of molecular lines, but here we concentrate on HDO which is the most relevant to the present study. Because the HDO transitions considered here require very high densities to be excited, we do not expect contamination from the off-signal despite the relatively short distance between the on- and off-positions.  

\subsection{Herschel/HIFI}

In our analysis, we included archived spectra from the {\sl Herschel} HIFI \citep{2010A&A...518L...6D} spectral scan towards OMC2-FIR4, performed in the course of the CHESS programme (PI C. Ceccarelli; \citealt{2010A&A...521L..22C}). The reduced HIFI Wide Band Spectrometer (WBS) spectra are available at the {\sl Herschel} Science Archive\footnote{https://www.cosmos.esa.int/web/herschel/hifi-spectral-scans-hpdp}. From these data, we have extracted lines of ortho- and para-$\water$, HDO, and high-$J$ CO. These spectra have previously been presented in \cite{2013A&A...556A..57K}, and some of them are shown in Figs.~\ref{ph2o_spectra}, \ref{oh2o_spectra}, and \ref{co_spectra} in the Appendix. HIFI observations were made in dual beam switching mode with the reference off-position set to $180\arcsec$ from the on-position. According to \citet[their Sect.~3.2]{2013A&A...556A..57K}, the spectra in question are not severely affected by contamination from the off-source signal and the self-absorption in the $\water$ spectra "appears to be related to the source".     

\section{Results}

\subsection{[OI]}
\label{OI}

The seven [\ion{O}{i}] $63\,\mu$m spectra observed with the upGREAT HFA array are shown in Fig.~\ref{o_spectra}. The actual positions of the pixels deviate from the regular hexagonal pattern shown in Fig.~\ref{hfa_on_pacs} by $\sim1-2\arcsec$. The centre position ($+0\farcs7,-1\farcs6$) shows a slightly asymmetric broad emission line that extends from $v_{\rm LSR} \sim -16\,\kms$ to $\sim 54\,\kms$. The integrated intensity of the line is $62.0\pm0.2\,{\rm K}\,\kms$ ($6.8\times10^{-6}\,{\rm W\,m^{-2}\,sr^{-1}}$), which translates into a total line flux of $F=7.2\times10^{-15}\,{\rm W\,m^{-2}}$, assuming that the emission comes from the solid angle spanned by the $6\farcs3$ beam. Adopting the distance 390\,pc, one obtains a line luminosity of $L({\rm \ion{O}{i}_{63\,\mu{m}}})=3.4\times10^{-2}\,L_\odot$. These numbers are comparable with previous, spectrally unresolved measurements with {\sl Herschel}/PACS for FIR4 reported by \citet[their Table~2]{2016A&A...596A..26G}. The broad emission line towards the centre peaks at an LSR velocity of $v_{\rm LSR}\sim10.8\,\kms$. The off-centre positions show mainly narrow lines, except in the north-east where the pixel encompasses {the Class II YSO MIPS 2297 and} part of the dense ridge seemingly connecting FIR4 to FIR3  (see Fig.~\ref{hfa_on_pacs}). Gaussian fits to the off-centre spectra give peak LSR velocities in the range $10.0-10.7\,\kms$ and line widths (FWHM) between $3.2\,\kms$ and $4.3\,\kms$. It seems natural to assume that the narrow lines originate in the PDR on the surface of the clump, whereas the broad emission line comes from the clump centre where shocks caused by protostellar outflows give rise to a large velocity dispersion. The narrow peak on top of the broad emission probably originates in the PDR. 

%OI
\begin{figure*}[htb]
\unitlength=1mm
\begin{picture}(160,135)(0,0)
\put(33,0){
\begin{picture}(0,0) 
\includegraphics[width=5.5cm,angle=0]{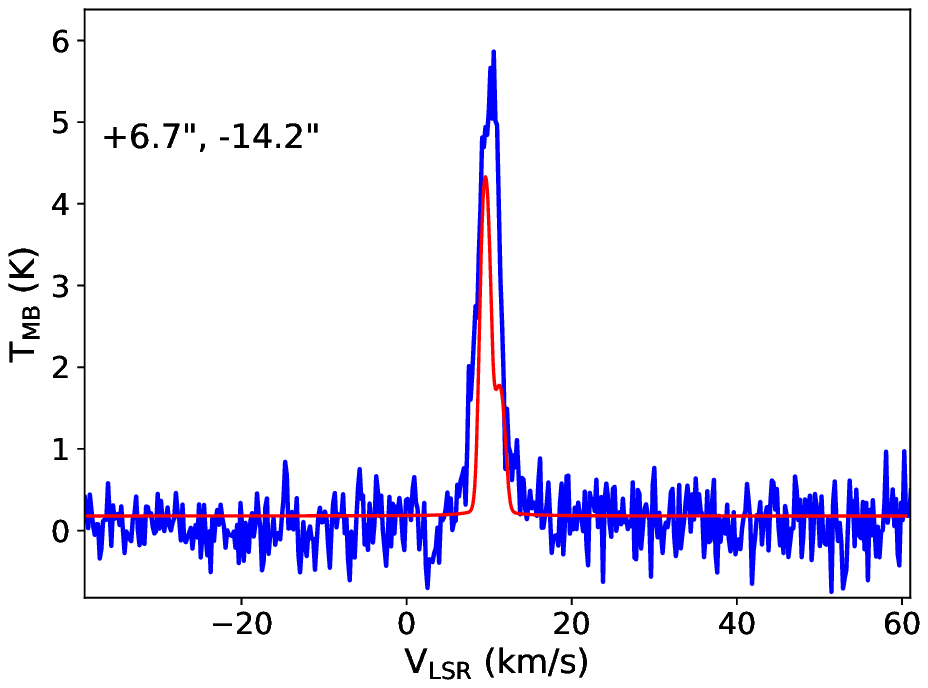}
\end{picture}}

\put(95,0){
\begin{picture}(0,0) 
\includegraphics[width=5.5cm,angle=0]{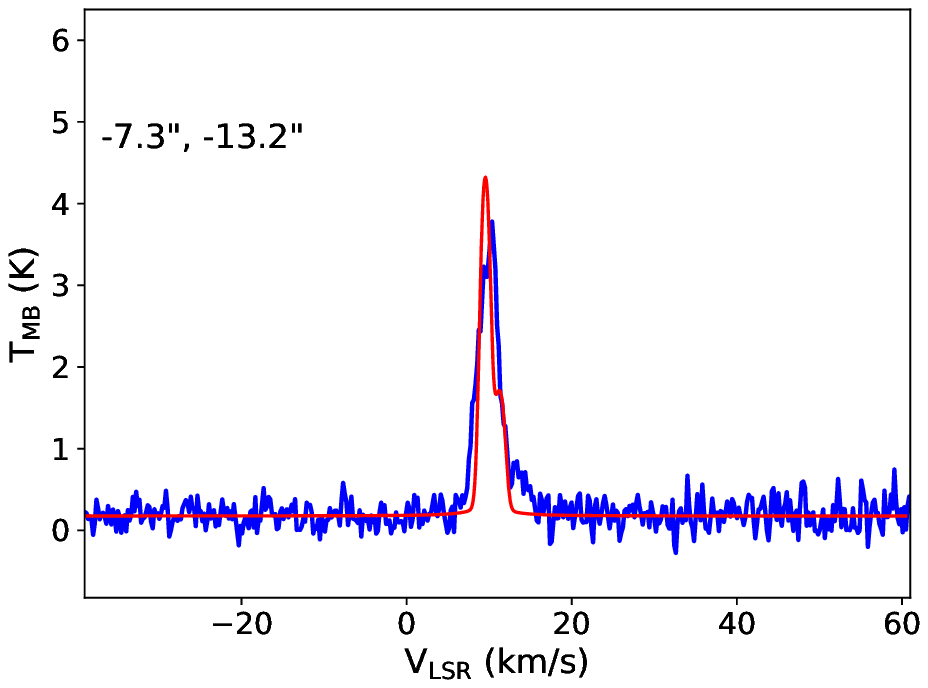}
\end{picture}}

\put(-5,50){
\begin{picture}(0,0) 
\includegraphics[width=5.5cm,angle=0]{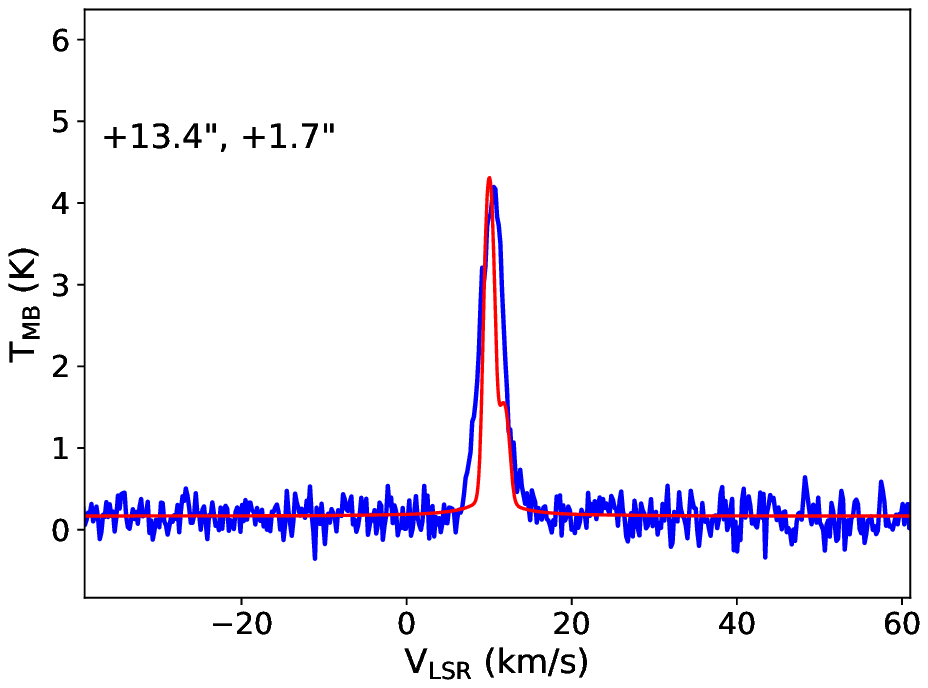}
\end{picture}}

\put(50,43){
\begin{picture}(0,0) 
\includegraphics[width=7cm,angle=0]{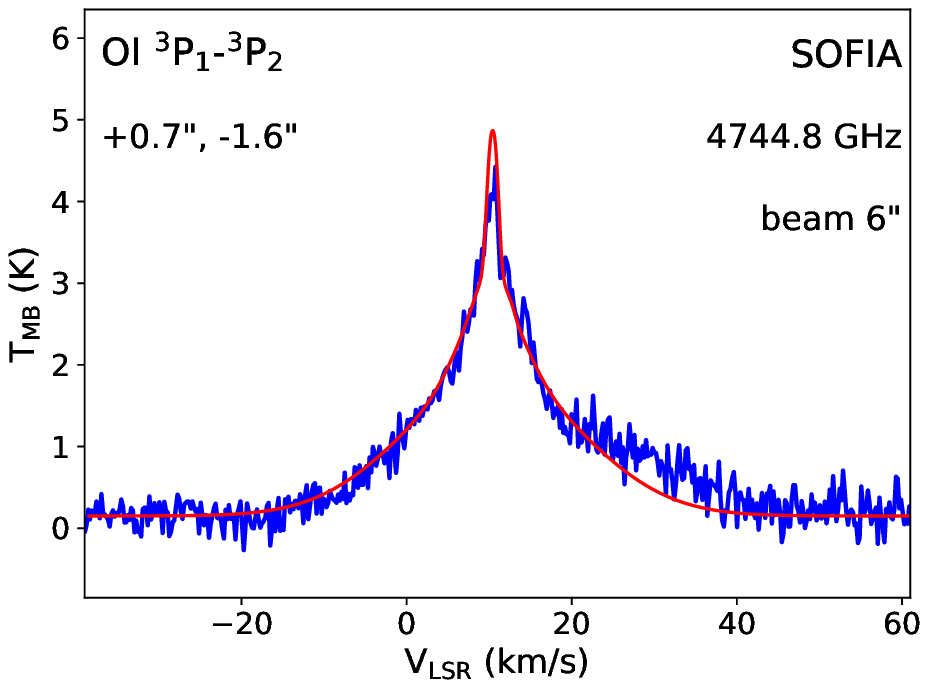}
\end{picture}}

\put(120,50){
\begin{picture}(0,0) 
\includegraphics[width=5.5cm,angle=0]{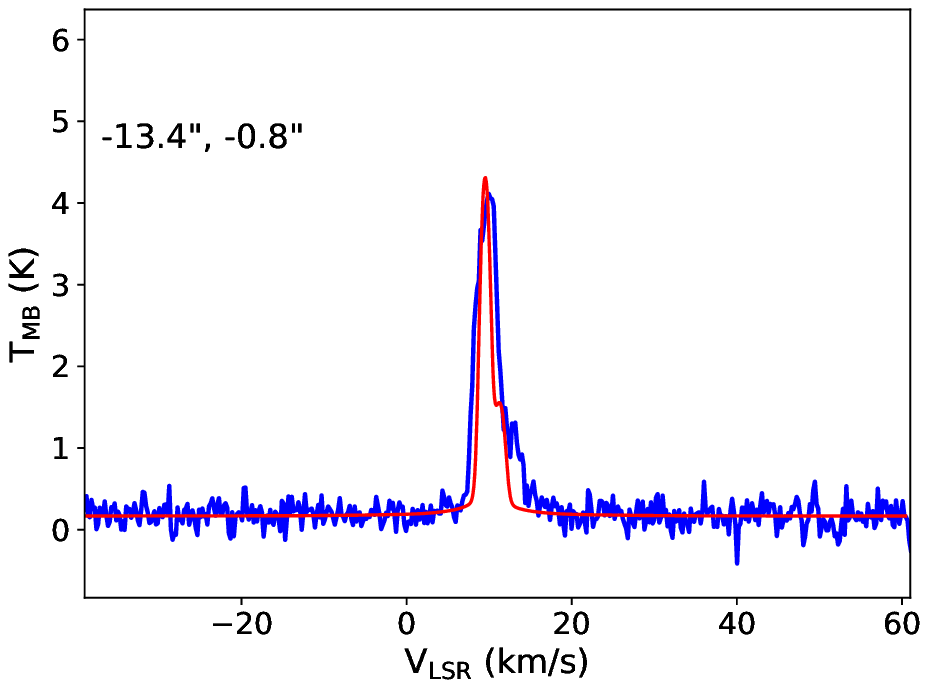}
\end{picture}}

\put(33,95){
\begin{picture}(0,0) 
\includegraphics[width=5.5cm,angle=0]{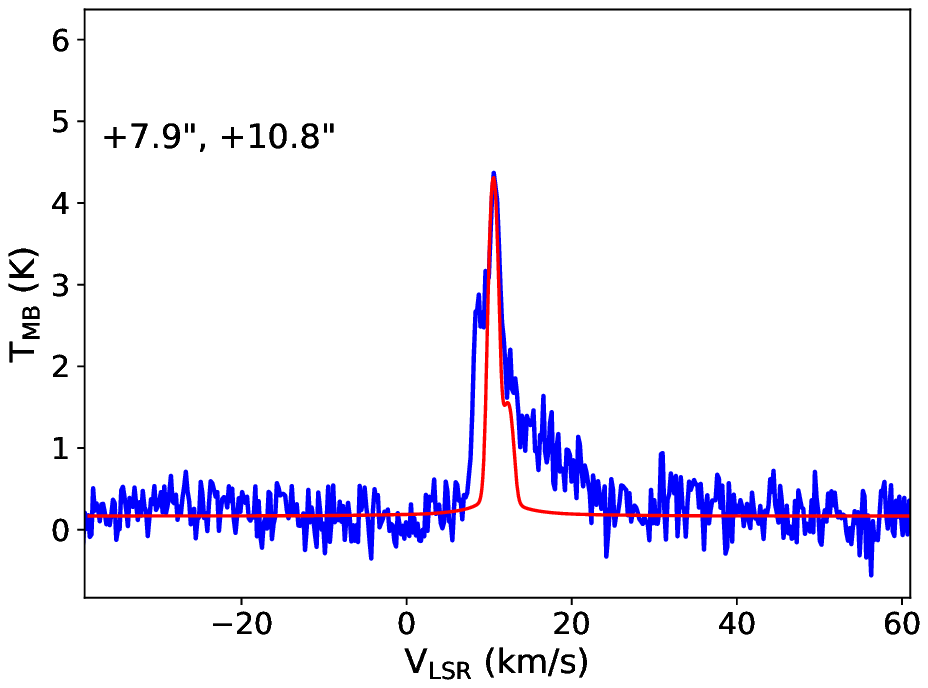}
\end{picture}}

\put(95,95){
\begin{picture}(0,0) 
\includegraphics[width=5.5cm,angle=0]{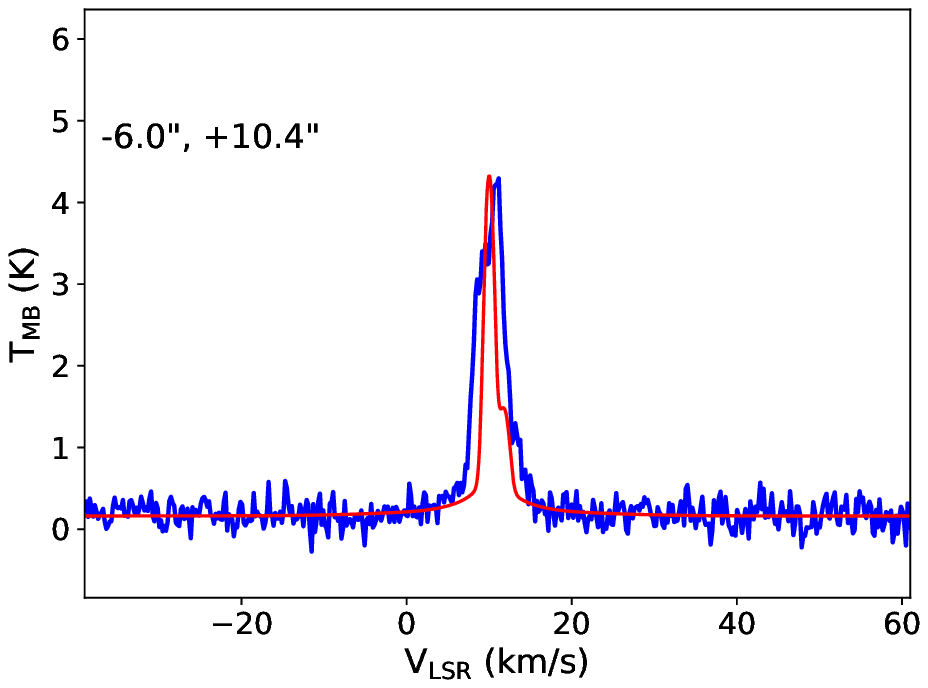}
\end{picture}}
\end{picture}
\caption{OI ${^3P}_1-{^3P}_2$ fine structure lines at 63\,$\mum$ observed with the upGREAT HFA array onboard SOFIA. The offsets from our (0,0) position, R.A. 05:35:27.0, Dec. -05:09:57, are indicated on the left of each panel. Spectra predicted from a spherical shell model described in Sect.~\ref{sect_model} are shown in red.}
\label{o_spectra}
\end{figure*}

\subsection{OH and HDO}

The $J=5/2^--3/2^+$ rotation line spectrum of OH in the electronic ground state $^2\Pi_{3/2}$ towards OMC-2 FIR4 is shown in Fig.~\ref{oh_spectrum}. In addition to a broad emission component, probably originating in the hot inner parts of the clump, the spectrum shows absorption by a cooler gas component in the foreground. The two strong hyperfine components of the line, $F,P=3,- \rightarrow 2,+$ (left) and $F,P=2,- \rightarrow 1,+$ (right), can be discerned in the absorption feature. The weaker third hyperfine component,  $F,P=2,- \rightarrow 2,+$, should appear on the right side of the two strong absorption lines in this spectrum. A hint of this component is seen in the model spectrum shown in red. A Gaussian fit to the hyperfine structure of the absorption feature gives an LSR velocity of $10.8\pm0.1\,\kms$ and a line width (FWHM) of $2.8\pm0.6\,\kms$.

\begin{figure}[ht]
\unitlength=1.0mm
\begin{picture}(80,60)(0,0)
\put(0,0){
\begin{picture}(0,0)
\includegraphics[width=8cm,angle=0]{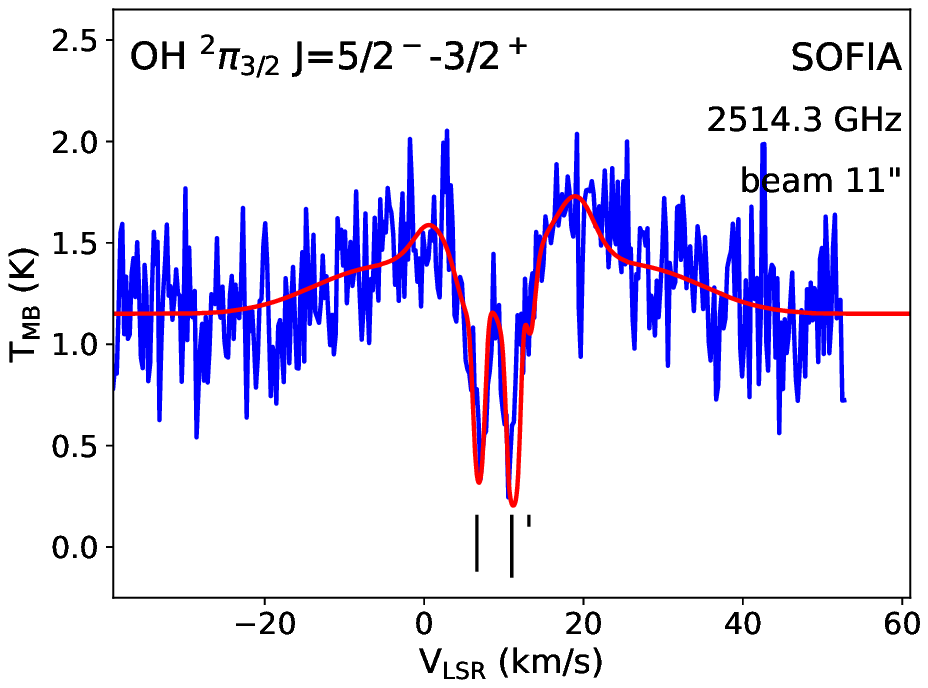}
\end{picture}}
\end{picture}
\caption{OH $^2\Pi_{3/2}(J=5/2^--3/2^+)$ spectrum at 2.51\,THz observed towards the centre of OMC-2 FIR4 with SOFIA/4GREAT-4. The spectrum predicted from the model described in Sect.~\ref{sect_model} is shown in red. The bars below the spectrum indicate the relative positions and line strengths of the three hyperfine components of the transition.}
\label{oh_spectrum}
\end{figure}

The HDO lines observed with SOFIA and APEX are shown in Fig.~\ref{hdo_spectra}. Here, we also include the $J_{K_a,K_b}=1_{11}-0_{00}$ line previously detected by {\sl Herschel}/HIFI. One can see that the non-detection of this line with SOFIA is consistent with the line intensity and the noise level obtained in the SOFIA spectrum. Of the two HDO lines observed with APEX only the ground-state line at 465\,GHz was detected. The HDO $1_{01}-0_{00}$ line shows a moderately broad pedestal and a narrow component that probably originates in dense, quiescent gas.

\begin{figure}[ht]
\unitlength=1.0mm
\begin{picture}(80,185)(0,0)

\put(10,135){
\begin{picture}(0,0)
\includegraphics[width=6.4cm,angle=0]{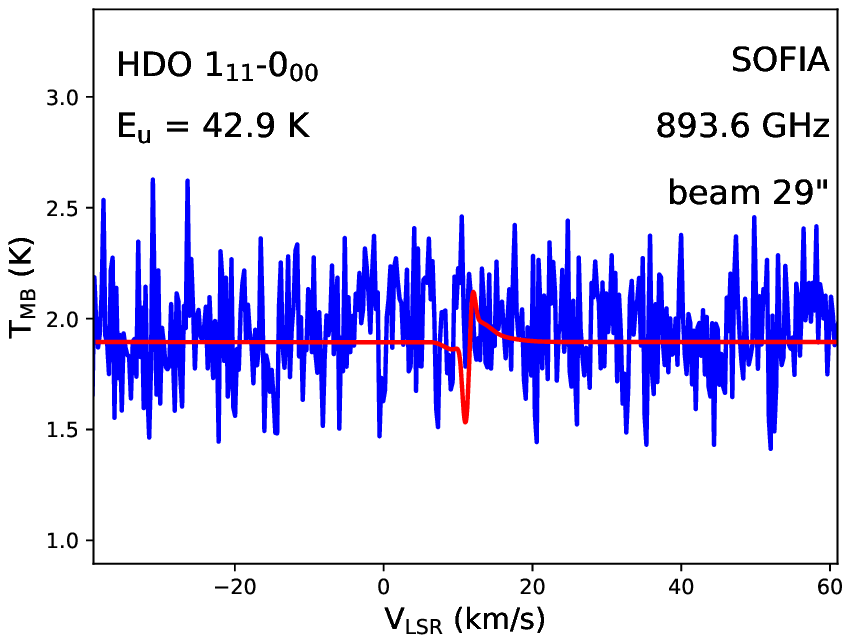}
\end{picture}}

\put(10,90){
\begin{picture}(0,0)
\includegraphics[width=6.4cm,angle=0]{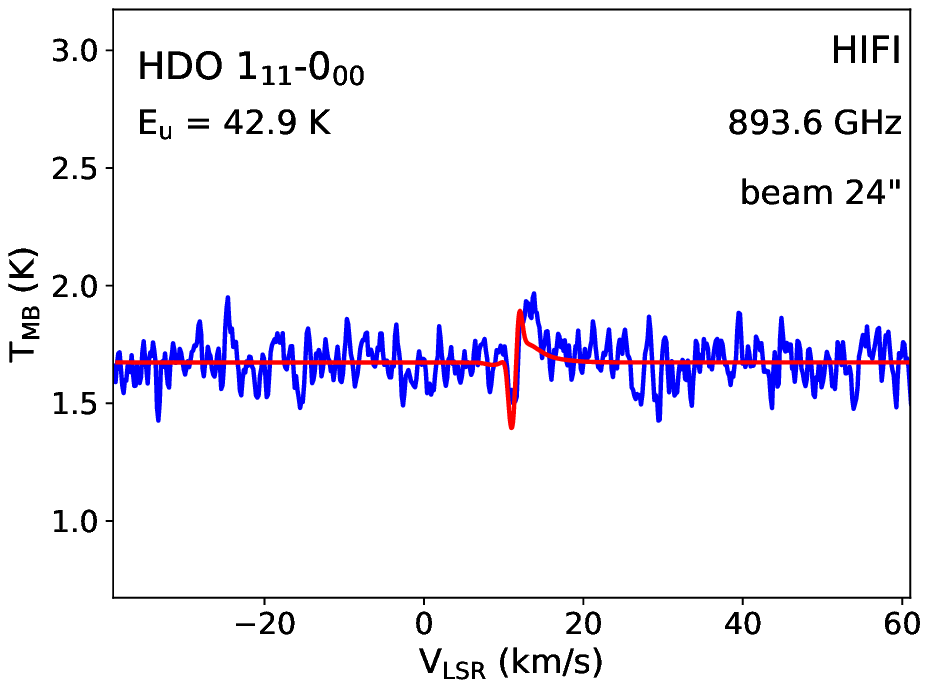}
\end{picture}}

\put(10,45){
\begin{picture}(0,0)
\includegraphics[width=6.4cm,angle=0]{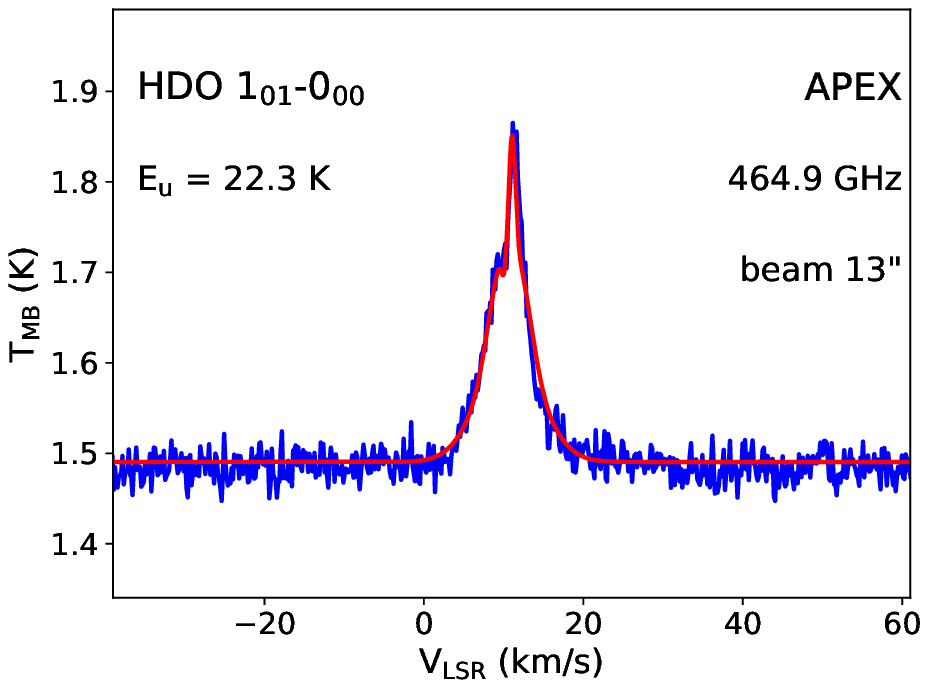} 
\end{picture}}

\put(10,0){
\begin{picture}(0,0)
\includegraphics[width=6.4cm,angle=0]{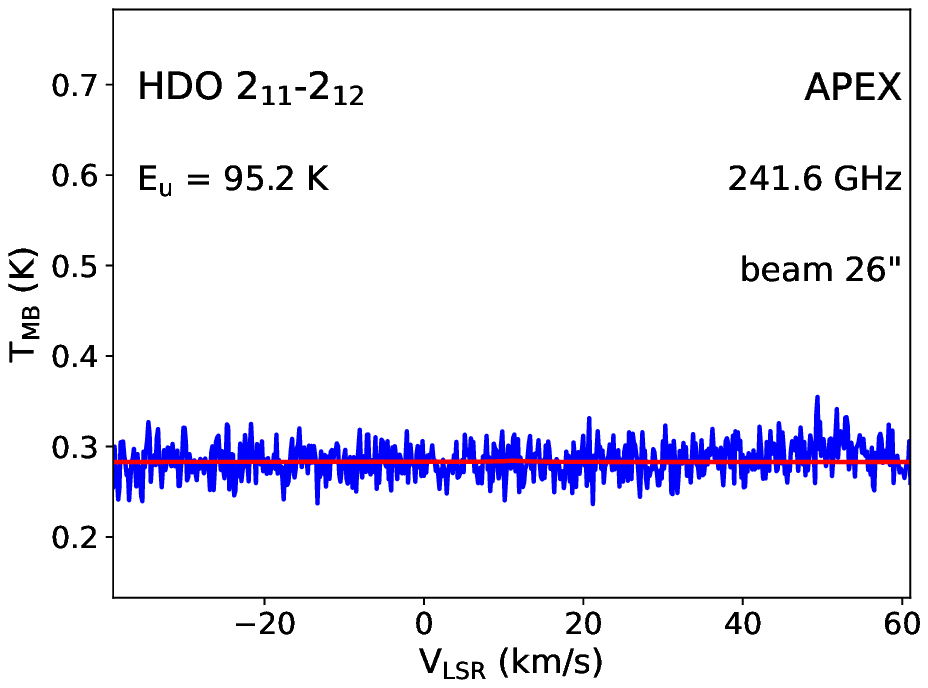}
\end{picture}}

\end{picture}
\caption{From top to bottom: HDO $1_{11}-0_{00}$ with SOFIA and {\sl Herschel}/HIFI, followed by $1_{01}-0_{00}$ and $2_{11}-2_{12}$ from APEX. Simulated spectra from the model described in Sect.~\ref{sect_model} are shown in red.}
\label{hdo_spectra}
\end{figure}

\section{Spherical shell model of OMC-2 FIR4}
\label{sect_model}

The spectral lines observed in the course of the present survey with SOFIA and APEX, as well as several other lines observed previously with {\sl Herschel}/HIFI, were simulated using a 1-D spherical model. The goal was to estimate the relative abundances of the main oxygen bearing species, \ion{O}{i}, OH, $\water$, and CO in the warm ($\sim 300-500$\,K) and cool (30-40\,K) components of the clump and in the PDR at its surface. In addition, we attempted to constrain the abundance of HDO in the shocked and cool gas components. 

The envelope around the protostar(s) was assumed to extend from a distance of $1000$\,au ($2\farcs6$) to 8\,000 au ($21\arcsec$) from the centre, and was assumed to be surrounded by a PDR extending to a radius of 12\,000 au ($31\arcsec$). The outer radius of the envelope corresponds approximately to the apparent radius of the clump in the $250\,\mum$ continuum, as observed with {\sl Herschel}/SPIRE. We tested different inner radii from 100\,au upwards. Values of the order of 1000\,au give the best overall agreement with the observations that we modelled and were performed using beam sizes ranging from $\sim 6\arcsec$ to $\sim 40\arcsec$. This size scale is consistent with the LVG analysis of the $\formyl$  and $\diaz$ lines from OMC-2 FIR4 by \cite{2014ApJ...790L...1C}, who found the best fit with the line intensities and ratios assuming a warm (120\,K) shell with a radius of $\sim 2000$\,au and a cool envelope (40\,K) with a radius of $\sim 4000$\,au. In addition, the  intensities of the lowest rotational lines of $\water$ observed by {\sl Herschel} indicate that the extent of the warm component is a few 1000\,au. This agrees with previous estimates for the sizes of "cavity shocks" associated with outflows from intermediate-mass and massive protostars (\citealt{2016A&A...585A.103S}; \citealt{2021A&A...648A..24V} and references therein). 

The density distribution was assumed to follow a power law $n\propto r^{-2}$ with a maximum $n(\htwo)=3\times10^7\,\percc$ at the inner boundary (1000\,au). The density at the outer edge of the envelope (8000\,au) was then $\sim 5\times10^5\,\percc$. The surrounding PDR was assumed to have a constant density of $3\times10^4\,\percc$. The dust temperature distribution was calculated taking into account the heating of the dust by both internal and external sources of radiation. The continuum radiative transfer program CRT \citep{2005A&A...440..531J} was used here. The external radiation field was assumed to be $10^4$ times stronger than the interstellar radiation field in the solar neighbourhood, as estimated by \cite{1982A&A...105..372M}. This corresponds to the estimate of \cite{2006A&A...449..609J} for OMC-2 FIR4 obtained from radiative transfer modelling of the spectral lines and the dust continuum.  

A 300\,K blackbody with a bolometric luminosity of $30\,L_\odot$ was placed at the centre of the spherical envelope. The central luminosity agrees with the estimates of \cite{2012ApJ...749L..24A} and \citet[$\sim30-50\,L_\odot$]{2014ApJ...786...26F}. The present dust model assumes "classical" grains with an MRN-like size distribution \citep{1977ApJ...217..425M} in radiative equilibrium. The gas temperature was assumed to be equal to the dust temperature inside the envelope but strongly elevated near its inner and outer edges. The temperature of the PDR shell needs to be near 100\,K to give rise to the strong, narrow \ion{O}{i} emission lines observed in off-centre pixels of the HFA. The warm outer layer is also consistent with the {\sl Herschel}/PACS detection of extended \ion{O}{i} emission around OMC-2 FIR4 reported in \cite{2016A&A...596A..26G}. Similarly, high gas temperatures in the inner regions were needed to reproduce the observed intensities of the broad \ion{O}{i}, OH, $\water$, and CO  lines. High gas temperatures can be envisioned to be caused by the photoelectric effect and gas-phase photo-processes, and, in the inner regions, also by mechanical shocks owing to protostellar outflows.

The non-thermal velocity distribution was assumed to be Gaussian, with the dispersion $\sigma_v$ (FWHM/$\sqrt{8 \ln 2}$) decreasing from $15\,\kms$ in the innermost shell to $0.5\,\kms$ in the outer layers. The blue-shifted self-absorption in the $\water$ spectra (Figs.~\ref{ph2o_spectra} and \ref{oh2o_spectra}) suggests that the inner envelope is expanding. Expansion speeds in the range $2-3\,\kms$ can reasonably well reproduce the shapes of the medium-broad component of the $\water$ lines and the slight asymmetries of the \ion{O}{i} and OH lines observed towards the centre of the clump. We fixed the expansion speed of the three innermost zones at $2\,\kms$. On the other hand, the shape of the 465 GHz HDO line (Fig.~\ref{hdo_spectra}) can be best produced by assuming a lower expansion speed of only $0.5\,\kms$ in the massive cool envelope. The \ion{O}{i} spectra in the off-centre pixels show a shoulder on the red side that can be approximately reproduced by assuming an expansion speed of $1\,\kms$.   

The envelope model with a power-law distribution of the density was divided into five zones, where the gas temperature, non-thermal velocity dispersion, the expansion speed, and the abundances of the observed species were set by hand. 
{The physical parameters of the zones and the fractional abundances of the molecules included in the modelling are listed in Table~\ref{zones}. Radial velocities, $V_{\rm rad}$, are given with respect to the centre of the clump. Here, they are always positive, meaning that the envelope is expanding. The density and temperature distributions of the model are visualised in Fig.~\ref{model}. The expansion speeds and Gaussian velocity dispersions $\sigma_v$ are also indicated in this graph.}

\begin{figure}
\unitlength=1mm
\begin{picture}(80,65)

\put(-5,0){
\begin{picture}(0,0) 
\includegraphics[width=9cm,angle=0]{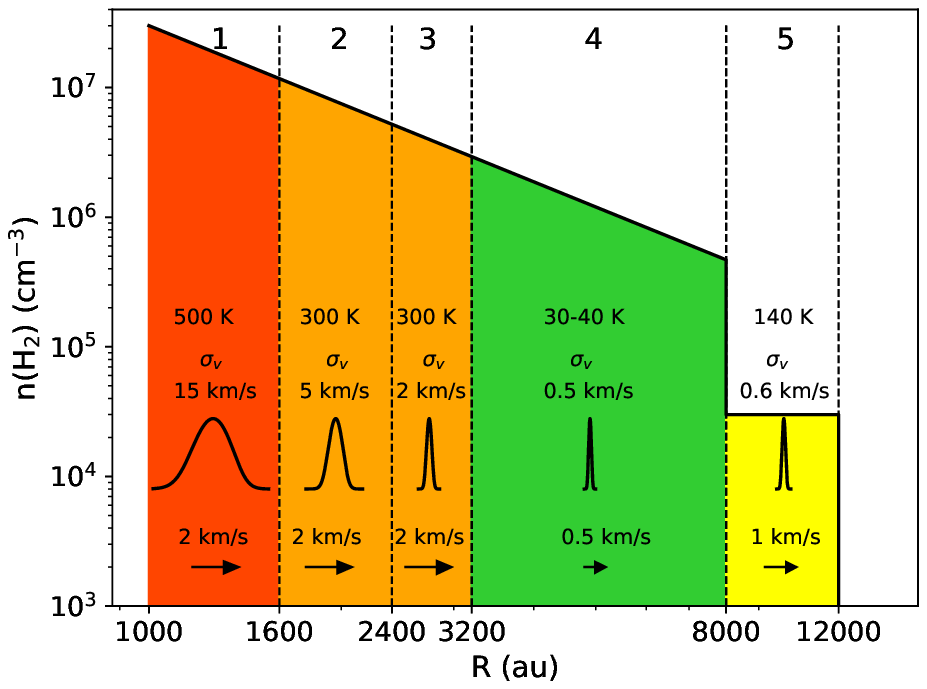}
\end{picture}}

\end{picture}
\caption{Visual presentation of the physical parameters of the spherical shell model with five zones. The number density is assumed to follow a power law $n(R) \propto R^{-2}$. The kinetic temperature in zone 4 (green) varies from $\sim 30$\,K to $\sim 40$\,K. It corresponds to the dust temperature in radiative equilibrium given the external and internal sources (see text). Elsewhere, the kinetic temperature is set by hand and is assumed to be constant within each zone. The arrows indicating the expansion speed and the width of the Gaussian profiles describing the non-thermal velocity dispersion are not on scale.}
\label{model}
\end{figure}

Spectral lines of observed species were simulated using the Line radiative transfer with OpenCL (LOC) program \citep{2020A&A...644A.151J} combined with the continuum radiative transfer program CRT \citep{2005A&A...440..531J}. With this combination, the simulations include the effects of dust on the observable lines: extinction by large dust columns, weakening of line intensities due to the FIR background generated by the dust, and the excitation of high-lying energy levels by this FIR radiation. The fractional abundances in the five zones were adjusted until an approximate agreement was reached between the simulated and observed line profiles. For simplicity, the fractional abundance of a species in a certain zone is non-zero only if this is necessary to explain the observed line profile. In the rest of the zones, the fractional abundance is set to zero either on chemical grounds or because the excitation conditions are such that the zone is likely to have little contribution to the observed line emission or absorption. Examples of the composition of the simulated spectra are shown in Fig.~\ref{decomposition}. According to the model, the narrow emission feature seen in the \ion{O}{i} spectra and the narrow absorption feature seen in the OH spectrum originate in the PDR (zone 5), while the narrow emission peak of the HDO $1_{01}-0_{00}$ spectrum comes from the cool part of the envelope (zone 4).     

\begin{table*}[htb]
\caption{Physical parameters and the fractional abundances in different zones of the spherical shell model of the envelope. }

\begin{tabular}{l|l|l|l|l|l} \hline
  \noalign{\smallskip}
  {zone}      &  \multicolumn{1}{c|}{1}        & \multicolumn{1}{c|}{2}         & \multicolumn{1}{c|}{3}         & \multicolumn{1}{c|}{4}         & \multicolumn{1}{c}{5} \\ 
  $R$ (au)  & \multicolumn{1}{c|}{1000-1600} & \multicolumn{1}{c|}{1600-2400} & \multicolumn{1}{c|}{2400-3200} & \multicolumn{1}{c|}{3200-8000} & \multicolumn{1}{c}{8000-12000} \\ \hline
  \noalign{\smallskip}
  $n(\htwo)$ ($\percc$) & $3\times10^7-10^7$ & $10^7-5\times10^6$  &  
  $5\times10^6-3\times10^6$ & $3\times10^6-5\times10^5$  &$3\times10^4$ \\
  $T$ (K)  &  500 & 300 & 300 & $30-40$ & 140  \\
  $V_{\rm rad}$ ($\kms$)  & 2 & 2 & 2 & 0.5 & 1 \\
  $\sigma_V$ ($\kms$)   &  15  & 5  & 2  & 0.5 & 0.6 \\ 
  $M/M_{\rm tot}^\dagger$ & 0.08 & 0.09 & 0.09 & 0.69 & 0.05 \\
  \noalign{\smallskip}
  \hline
  \noalign{\smallskip}
  %\rowcolor{white}
   %\multicolumn{7}{c}{} \\ 
  $X({\rm \ion{O}{i}})^*$     & $3\times10^{-6}$ & $6\times10^{-7}$ & $1\times10^{-7}$ & - & $2\times10^{-4}$ \\
  $X({\rm OH})$     & $1.1\times10^{-8}$ & $2.3\times10^{-8}$ & $5\times10^{-9}$ & - & $9\times10^{-8}$ \\
  $X({\rm HDO})$    & - & $3.5\times10^{-11}$ & $3.5\times10^{-11}$ & $4\times10^{-11}$ &  - \\ 
   $X({\rm o\water})$ & $2.7\times10^{-9}$ & $9\times10^{-9}$ & $2\times10^{-8}$ & - & $4\times 10^{-8}$ \\   
  $X({\rm p\water})$ & $1.8\times10^{-9}$ & $4\times10^{-9}$ & $6\times10^{-9}$ & - & $1.3\times10^{-8}$ \\  
 
    $X({\rm CO})^\ddagger$ & $8\times10^{-7}$ & $2\times10^{-6}$ & $5\times10^{-7}$  & - & - \\ 
    \noalign{\smallskip}
    \hline 
 \noalign{\smallskip}
\end{tabular}

$^\dagger$ The fraction of mass in the envelope with a total mass of $23\,M_\odot$.

$^*$ A dash means that the abundance is assumed to be negligible in that zone, or that on grounds of excitation conditions, the observed line is insensitive to the abundance in that shell.

$\ddagger$ Only the HIFI spectra are used for this fit.   

\label{zones}
\end{table*}
%\end{center}

  The model presented here is an idealization that in some average sense describes the sizes, densities, and temperatures of the different components of the envelope. This is suggested by the fact that the model can reasonably well reproduce the shapes and intensities of several spectral lines observed towards this source. We therefore believe that the abundance ratios given by this model are correct within an order of magnitude.  

The simulated lines of \ion{O}{i}, OH, and HDO are shown in red in Figs.~\ref{o_spectra}, \ref{oh_spectrum}, and \ref{hdo_spectra} superposed on the observed spectra. In Figs.~\ref{ph2o_spectra}, \ref{oh2o_spectra}, and \ref{co_spectra} in the Appendix we show the lowest rotational lines of para-$\water$ and ortho-$\water$ observed with {\sl Herschel}/HIFI, together with three high-$J$ CO lines that are supposed to probe the same  gas component as the high-velocity water lines \citep{2021A&A...648A..24V}. In these spectral figures, the lines predicted by the present spherical shell model are also shown in red. 

\section{Discussion}

\subsection{PDR}

The present model of OMC-2 FIR4 which consists of five shells was invoked in order to estimate the relative abundances of the major oxygen bearing species near the surface and in the inner, shocked parts of the clump.  The outermost shell represents the PDR that has previously been found to sheathe the molecular ridge of OMC-2 (\citealt{1997ApJ...481..343H}; \citealt{2016A&A...596A..26G}). Assuming constant abundances in this PDR as we have done is a simplification. According to the models of \cite{2009ApJ...690.1497H}, the relative abundances of \ion{O}{i}, OH, $\water$, and CO change rapidly in a PDR as functions of the visual extinction $A_{\rm V}$ calculated from the cloud surface (see their Figure 3). This depends on photodissociation near the surface, accretion of atomic oxygen on the grain surfaces, and subsequent formation of water ice and its photo-desorption. In the models of \cite{2009ApJ...690.1497H} the gas-phase abundances of $\water$, OH, O$_2$, and CO reach a maximum at depths $A_{\rm V}\sim$ a few magnitudes into the cloud, depending on the external FUV field, and remain almost constant for another $\Delta A_{\rm V}\sim$ a few magnitudes, before freezing out.  The thickness of the outermost shell in our model corresponds to a hydrogen column density of $N_{\rm H} \sim 3.6\times10^{21}\,\persqcm$ or a visual extinction of $A_{\rm V}\sim 3.6^{\rm mag}$. However, the clump is embedded in an ambient cloud, so the effective visual extinction at the bottom of this layer is probably higher than that. Using the analytical formula of \cite{2009ApJ...690.1497H} for the plateau where these abundances peak (their Eqs. (20), (31), and (32)), the relative abundances of O, OH, and $\water$ estimated in the outer layer, $\sim 2200:1.7:1$ (see Table~\ref{zones}),  correspond to a visual extinction of $A_{\rm V}\sim 4.2^{\rm mag}$ (assuming $G_0=10^4$ and $\sigma_{\rm H}=5\times10^{-22}\,{\rm cm^2}$).  

{The bright elongated \ion{O}{i} feature seen in the spectrally unresolved map from  {\sl Herschel}/PACS \citep{2016A&A...596A..26G} coincides with the dense ridge connecting OMC-2 FIR3 and FIR4 (Fig.~\ref{hfa_on_pacs}) but is also aligned with a radio jet and a powerful CO outflow that seem to emanate from HOPS 370 in FIR3 (\citealt{2017ApJ...840...36O}; \citealt{2023ApJ...944...92S}). \cite{2016A&A...596A..26G} argued that the \ion{O}{i} feature can be attributed to the jet from HOPS 370 and that the \ion{O}{i} emission peak towards FIR4 represents a terminal shock of this jet. Based on the resolved spectra shown in Fig.~\ref{o_spectra} and recent interferometric observations discussed below, we think that the intensity maximum towards FIR4 can be more naturally explained by outflows that have their origins inside the FIR4 clump. In addition, we suggest that the dense PDR surrounding OMC-2 that produces the emission detected in the off-centre pixels of the HFA (Fig.~\ref{o_spectra}) can have a substantial contribution to the elongated \ion{O}{i} feature between FIR3 and FIR4. This region can be relatively bright in \ion{O}{i} owing to the high column densities. The fine-structure line of singly ionised carbon, \ion{C}{ii}, at $158\,\mum$ is another tracer of PDRs, although it probes their less dense outer parts (\citealt{1999ApJ...527..795K}; see also Figure 4 in \citealt{2009ApJ...690.1497H}). The OMC-2 cloud and the ridge between FIR3 and FIR4 do not stand out in the large-scale [\ion{C}{ii}] mapping of Orion A with SOFIA presented in \cite{2021A&A...651A.111P} and \cite{2021A&A...652A..77H}. The integrated [\ion{C}{ii}] line intensity map is dominated by limb-brightened shells associated with M42, M43, and NGC 1977 and an extended faint component. A comparison of the [\ion{C}{ii}] and [\ion{O}{i}] spectra towards the positions observed in the present survey is shown in Fig.~\ref{cii_oi_spectra}. The LSR velocity and width of the narrow [\ion{O}{i}] component are similar to those of [\ion{C}{i}]. The ratio of the integrated intensities [\ion{O}{i}]/[\ion{C}{ii}] ranges from $\sim1.5$ to $\sim 2.8$, expect for our (0,0) ($\sim 9.1$) and the offset ($+7\farcs9,+10\farcs8$) ($\sim3.6$) where the [\ion{O}{i}] line has a red-shifted wing. These ratios are in agreement with the predictions for a dense PDR with a strong incident FUV flux \citep{1999ApJ...527..795K}. The [\ion{O}{i}] line that originates in a jet would probably show velocity shifts and/or large line widths. Spectrally resolved \ion{O}{I} mapping of the ridge is needed to confirm the presence of the putative \ion{O}{i} jet.}  

\subsection{Warm inner envelope}

The emission of atomic oxygen associated with protostars can originate in wide-angle winds or collimated jets launched by protostar-disk systems, terminal or internal shocks within jets, or shocks occurring when the wide-angle winds collide with the molecular envelope surrounding the protostar (\citealt{1985Icar...61...36H}; \citealt{2010A&A...518L.121V}; \citealt{2015ApJ...801..121N}; \citealt{2017A&A...601L...4K}; \citealt{2017A&A...602A...8G}; \citealt{2021A&A...650A.173S}). The different structural components have kinematic characteristics that may be recognized in resolved \ion{O}{i} spectra. The present \ion{O}{i} spectra with the detection range extending from $v_{\rm LSR} \sim -50$ to  $\sim +100\,\kms$ show no evidence of very high velocity jets or bullets. Similarly, the CO $5-4$ spectrum observed with SOFIA is clean of line emission outside the velocity range shown in Fig.~\ref{co_spectra} up to $v_{\rm LSR}\sim\pm200\,\kms$. Judging from the size of the emission region and its physical conditions inferred from the modelling we assume that the observed broad line emissions of \ion{O}{i}, OH and $\water$ come from outflow cavity shocks or jet terminal shocks in the inner parts of the FIR4 clump. 
 
The \ion{O}{i} spectrum towards OMC-2 FIR4 (excluding the PDR component) is made up of a broad ($\sigma_v=15\,\kms$), nearly Gaussian component, and two narrower components ($\sigma_v=5\,\kms$ and $\sigma_v=2\,\kms$), which are also seen in the OH, $\water$ and high-$J$ CO spectra towards this source. The spectrum is measured towards the most prominent protostars, HOPS-108 and HOPS-64, embedded in the clump. The broadest line component corresponds to the innermost zone of our model (Table~\ref{zones}), where the abundance of \ion{O}{i} is approximately 700 times that of (ortho+para) $\water$ and 300 times that of OH. The high relative abundance of atomic oxygen suggests that the broad component originates in the cooling gas behind the fast molecular shocks that have been modelled by \cite{1989ApJ...340..869N}, \cite{1989ApJ...342..306H}, and, more recently, by \cite{2020A&A...643A.101L}. According to these models, shocks moving at speeds exceeding the ion magnetosonic speed of the medium (typically above $\sim 30\,\kms$) generate strong UV radiation that ionises and dissociates atoms and molecules in both the pre-shock and post-shock gas components, and heats these through the photoelectric effect. The component responsible for the broad \ion{O}{i} emission may correspond to the second temperature plateau of a dissociative J-shock, where the re-formation of $\htwo$ on the grains maintains the temperature at $\sim 400-500$\,K (see Fig.~14 of \citealt{1989ApJ...340..869N}). The plateau appears at high pre-shock densities ($n>10^5\,\percc$), where freshly formed $\htwo$ in excited vibrational states is collisionally de-excited by atomic hydrogen. This transforms the vibrational energy to the kinetic energy of the atoms and delays radiative cooling \citep{1989ApJ...342..306H}. 

In lower-velocity C-shocks, where neutrals and the ion-electron fluid have different flow velocities and temperatures, and the impact is mediated by ion-neutral collisions, $\htwo$ is not significantly dissociated. In these conditions, $\water$ and CO are supposed to become the most important coolants (\citealt{1996ApJ...456..250K}; \citealt{2019A&A...622A.100G}; \citealt{2023A&A...675A..86K}). Low-velocity shocks (with speeds around $\sim 5\,\kms$) where the gas temperature does not rise high enough for efficient $\water$ formation are an exception, where the abundance of atomic O is predicted to be much higher than those of $\water$ and CO (see, for example, Table~2 in \citealt{2019A&A...622A.100G}).  

The two model zones in the inner envelope with $\sigma_v=5\,\kms$ (zone 2) and $\sigma_v=2\,\kms$ (zone 3) have higher fractional abundances of $\water$ than zone 1, although remaining below those of \ion{O}{i} and CO. The \ion{O}{i}/CO ratio decreases from $\sim4$ in zone 1 to below 1 in zones 2 and 3. These tendencies indicate an increase in the proportion of molecular gas in zones 2 and 3. These zones may represent post-shock gas behind irradiated shocks or gas adjacent to shock fronts and heated by radiative or magnetic precursors, where water is produced in high-temperature gas-phase reactions or released from grains (and is also destroyed by photodissociation)  \citep{2008A&A...482..549J}. The latter alternative is supported by the fact that the HDO/$\water$ ratio derived in these zones is $\sim 0.001 -0.003$ (Table~\ref{zones}), which corresponds to values measured in massive hot cores, where the ices have been completely sublimated (\citealt{2021A&A...648A..24V} and references therein; \citealt{2024A&A...688A..29S}).  According to Table~\ref{zones}, the $\water/{\rm OH}$ ratio increases strongly from zone 2 ($\sigma_v=5\,\kms$) to zone 3 ($\sigma_v=2\,\kms$), suggesting a weakening of the radiation field.  

The physical conditions in an envelope are likely to change gradually with distance along and from the outflow axis (see, e.g., Fig.~8 in \citealt{2016A&A...585A.103S}), so the geometry around a single outflow is cylindrical or conical rather than spherical. The OMC-2 FIR4 clump contains several jet-outflow systems with different orientations. The protostars HOPS-108 (Class 0) and HOPS-64 (Class I), and the millimetre source "FraSCO 33" \citep{2023ApJ...944...92S} are located within $3\arcsec$ of the observed position and are therefore encompassed in the telescope beam for all the lines observed here. According to \cite{2023ApJ...944...92S}, the three objects are associated with their own compact but rather disordered CO outflows: "flow-6", "flow-4", and "flow-5" (see Figure 10 in \citealt{2023ApJ...944...92S}). The outflow associated with HOPS-108, "flow-6",  has the largest velocity range, extending from $-5\,\kms$ to $\sim 50\,\kms$. The collimated SiO jet discussed by \cite{2023A&A...671A..35L} extends to the south-east from the vicinity of HOPS-108 and may therefore be related to "flow-6".  It is probably identifiable with the prominent SiO emission structure labeled "shock 7" by \cite{2023ApJ...944...92S}, which according to these authors is, however, caused by the energetic outflow from the intermediate-mass protostar HOPS-370 located in the FIR3 clump. The brightest and most extensive region of SiO emission (called "shock-6" by \citealt{2023ApJ...944...92S}) is found on the western side of HOPS-108 and HOPS-64 in the same area as the compact outflows "flow-4" and "flow-5". However, \cite{2023ApJ...944...92S} favour the scenario presented previously by \cite{2008ApJ...683..255S} that this extensive shock is also caused by the HOPS-370 outflow.    

 Although evidence for a strong influence from the HOPS-370 outflow on the FIR4 clump seems solid, we consider it probable that the rather symmetric high-velocity wings seen in the \ion{O}{i}, CO, and $\water$ lines towards the centre of the clump arise from shocks induced by internal sources. In accordance with the conclusions of \cite{2010A&A...518L.121V} in the case of the low-mass protostar HH-46, we suggest that the broad \ion{O}{i} line originates in high-velocity J-shocks near the embedded YSOs. Then it would be  natural to assume that the wings of the $\water$ and CO lines come from the same gas. Judging from the low $\water/{\rm \ion{O}{i}}$ and $\water/$CO ratios, the narrower line components also come from dissociative J-shocks or externally irradiated C-shocks. In the former case, the narrow components could represent cooler gas further downstream of the broad component, which is the brightest in \ion{O}{i}. As to the origin of the line wings, these ideas diverge from the concept presented in several previous studies of low-mass star-forming regions where the broad Gaussian component seen in the $\water$ and high-$J$ CO spectra is attributed to non-dissociative shocks along the outflow cavities, whereas "medium broad" ($\Delta v < 10\,\kms$) components are explained by jet-induced dissociative "spot shocks" (\citealt{2013A&A...557A..23K}; \citealt{2014A&A...572A..21M}; \citealt{2021A&A...648A..24V}).

\section{Conclusions}

Spectrally resolved observations with the High Frequency Array (HFA) of the upGREAT receiver onboard SOFIA show that the strong \ion{O}{i} emission at 63\,$\mu$m towards the protostellar clump OMC-2 FIR4 is dominated by broad wings that are similar to those seen in OH, $\water$ and high-$J$ CO lines towards this source. We assume that these lines originate in shocks caused by protostars embedded in the clump. Using an expanding spherical shell model that reproduces the observed lines reasonably well, we infer that oxygen is mainly atomic in the broad wing component. This component  probably represents the post-shock cooling gas behind dissociative J-shocks that generate strong UV radiation as described in \cite{1989ApJ...340..869N}, \cite{1989ApJ...342..306H}, and \cite{2020A&A...643A.101L}. According to our shell model, the components with medium-wide lines ($\sigma_v  \leq 5\, \kms$) still have more \ion{O}{i} than $\water$ but the proportion of molecular gas is greater; the abundance ratio \ion{O}{i}/CO is less than unity there. The HDO/$\water$ abundance ratio in this component is of the order of $10^{-3}$, which is supposed to represent the average deuterium fractionation ratio in the icy mantles of dust grains. Therefore, it is not necessary to invoke high-temperature chemistry in addition to ice sublimation to explain the presence of gaseous water in the component with medium-wide lines.   

{The six off-centre pixels of the upGREAT/HFA receiver show narrow $63\,\mum$ \ion{O}{i} lines that are likely to originate in the extended PDR at the surface of the OMC-2 cloud (\citealt{1997ApJ...481..343H}; \citealt{2016A&A...596A..26G}). We suggest that the elongated \ion{O}{i} feature seen previously in the {\sl Herschel}/PACS \ion{O}{i} map between FIR3 and FIR4 probably has a contribution from the PDR in the direction of the dense ridge connecting the two clumps.}      

\begin{acknowledgements}
The authors thank the SOFIA and APEX telescope staff for performing the observations, Mika Juvela for helpful discussions, {the anonymous referee for comments that helped to improve the manuscript}, and the Max Planck Society for financial support. 

\end{acknowledgements}

\bibliographystyle{aa} % style aa.bst

\bibliography{bibliography.bib} 

\begin{appendix}

\section{Oxygen spectra towards the reference positions}

The off-phase signals recovered in the post-flight processing of the \ion{O}{i} spectra are shown in Fig.~\ref{off_spectra}. These are averages of the seven HFA pixels. The positions lie $\pm 250\arcsec$ east and west of our (0,0), R.A. 05:35:27.0, Dec. -05:09:57 (J2000.0). The parameters of Gaussian fits to the line profiles are given in Table~\ref{off_gaussian}.

\begin{table}[htb]
\caption{Gaussian parameters of the averaged \ion{O}{i} spectra from the off positions.}

\begin{tabular}{cccc} \hline
  \noalign{\smallskip}
   & $T_{\rm MB}$ & $v_{\rm LSR}$ & $\Delta v$  \\
   & (K) & ($\kms$) & ($\kms$) \\ \hline
  \noalign{\smallskip}
  OFF A & 1.48 (0.18) & 11.73 (0.07) & 4.22 (0.16) \\
  OFF B & 1.68 (0.20) & 10.56 (0.06) & 3.60 (0.17) \\ 
      \noalign{\smallskip}
    \hline 
 \noalign{\smallskip}
\end{tabular}
\label{off_gaussian}
\end{table}

\begin{figure}[ht]
\unitlength=1.0mm
\begin{picture}(80,60)(0,0)
\put(0,0){
\begin{picture}(0,0)
\includegraphics[width=8cm,angle=0]{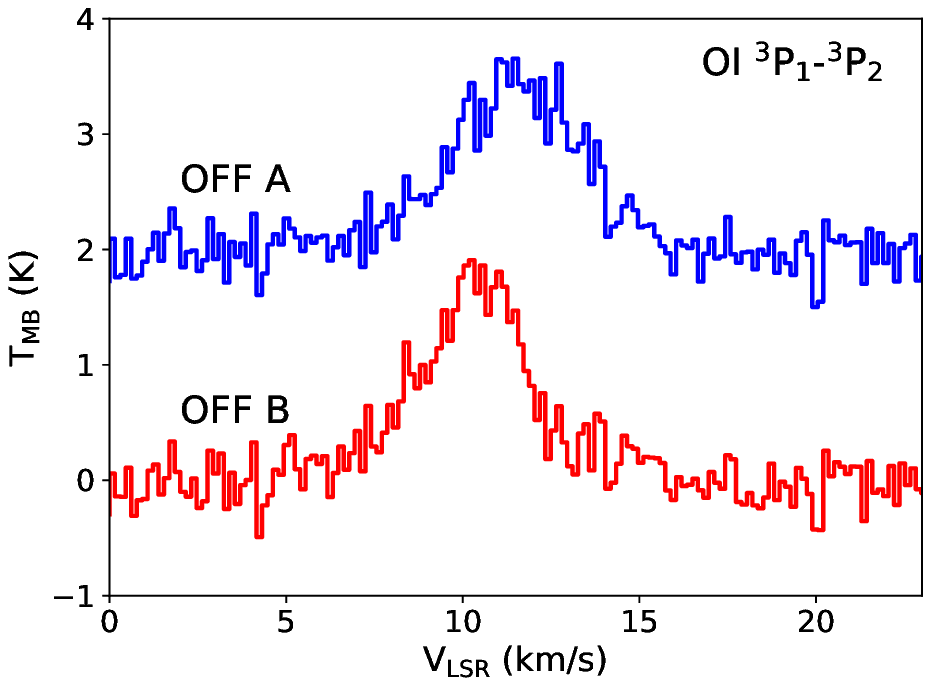}
\end{picture}}
\end{picture}
\caption{\ion{O}{i} spectra towards the two reference positions used in the dual beam switching observations. "OFF A" lies $250\arcsec$ east (positive R.A.) of our (0,0) and "OFF B" lies $250\arcsec$ west of it. The "OFF A" spectrum is shifted up by 2\,K for display purposes.}
\label{off_spectra}
\end{figure}

\section{Observed and simulated {\sl Herschel}/HIFI spectra}

The four lowest para- and ortho-$\water$ lines observed towards OMC-2 FIR4 with {\sl Herschel}/HIFI are shown in Figs.~\ref{ph2o_spectra} and \ref{oh2o_spectra}. The simulated spectra using the physical model and the fractional abundances given in Table~\ref{zones} are shown as red curves. The CO $J=5-4$ spectrum observed with SOFIA is shown in Fig.~\ref{co_spectra} (top left panel). The other panels of this figure show the CO $J=13-12$, $14-13$, and $15-14$ lines observed with {\sl Herschel}/HIFI. The simulated spectra are {coloured} red. One can see that the model that approximately reproduces the high-$J$ CO lines does not work for the CO $5-4$ line that was observed with a beam of $55\arcsec$. As discussed in \cite{2013A&A...557A..23K} and \cite{2014A&A...572A..21M}, low-$J$ ($\leq 6-5$) CO lines trace cool, entrained outflow gas on large scales, whereas high-$J$ lines come from fresh shocked gas, the same component that also gives rise to the bright $\water$ lines. However, by adjusting the CO abundances in the inner and the outermost zones of the model, a profile can be achieved that resembles the observed CO 5-4 line (dashed curve in the top left panel). The narrow emission feature seen in the CO 5-4 line probably originates in the PDR (zone 5 of the model). The high-$J$ CO lines are not excited in the cool quiescent gas and the PDR (zones 4 and 5), and the CO abundance there does not affect these profiles.  

%p-H2O
\begin{figure*}[htb]
\unitlength=1mm
\begin{picture}(160,110)(0,0)

\put(5,55){
\begin{picture}(0,0) 
  \includegraphics[width=7.3cm,angle=0]{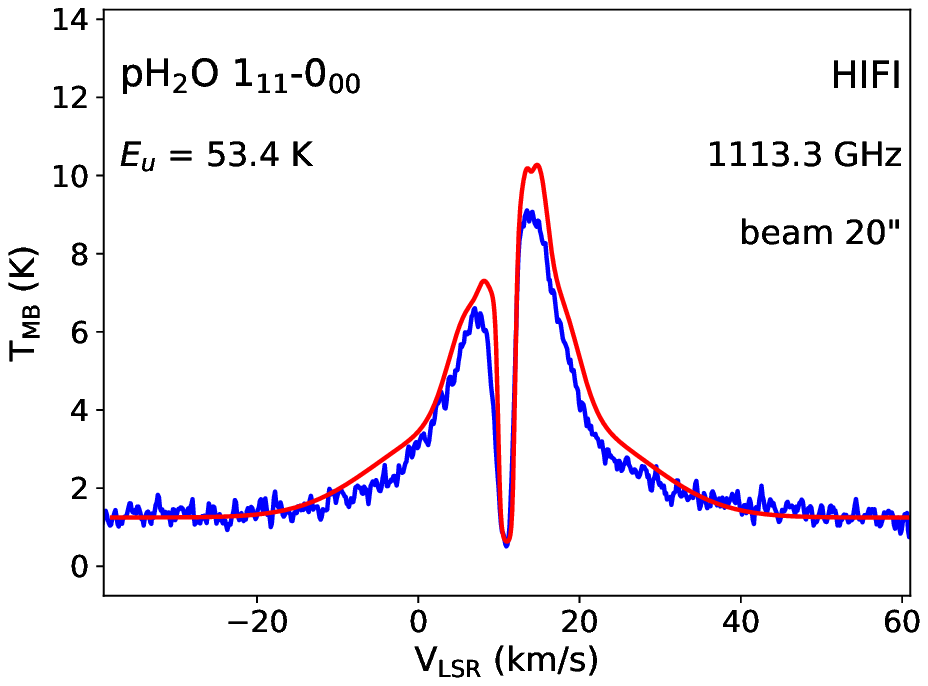}
\end{picture}}  
\put(90,55){
\begin{picture}(0,0) 
  \includegraphics[width=7.3cm,angle=0]{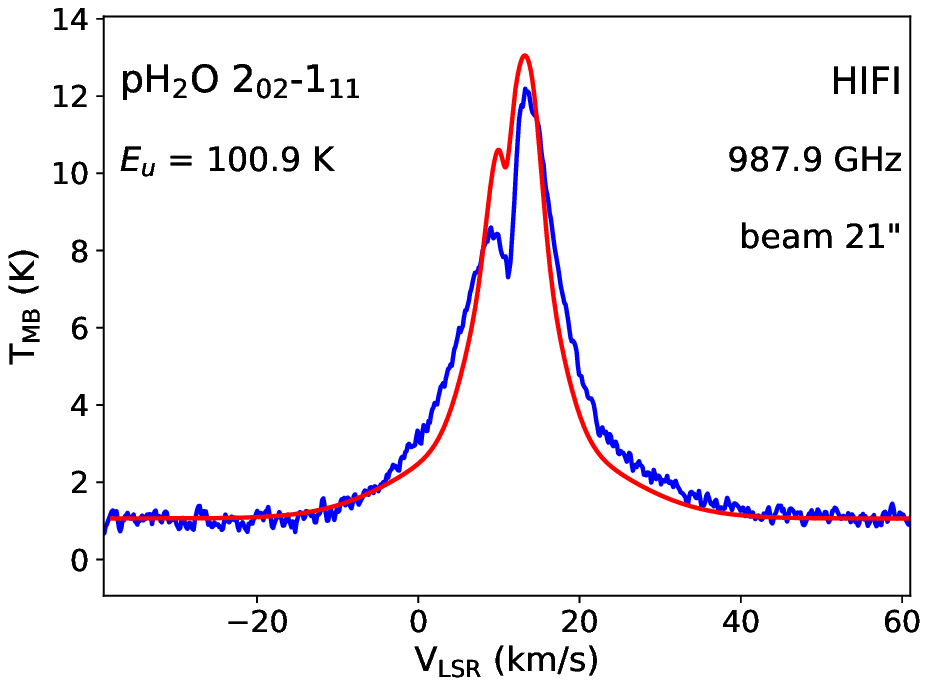}
\end{picture}}  
\put(5,0){
\begin{picture}(0,0) 
  \includegraphics[width=7.3cm,angle=0]{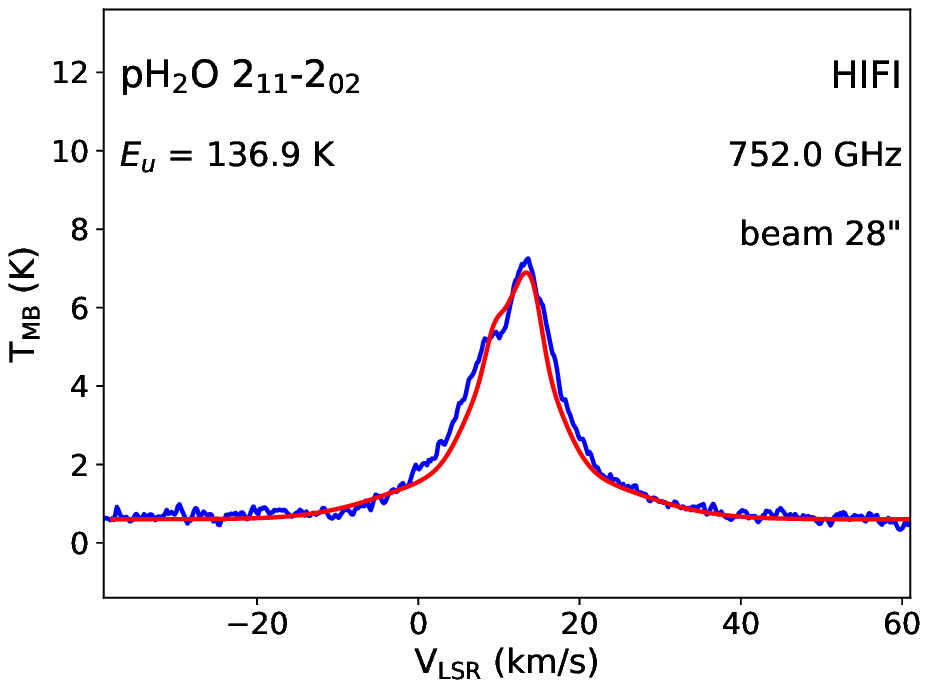}
\end{picture}}  
\put(90,0){
\begin{picture}(0,0) 
  \includegraphics[width=7.3cm,angle=0]{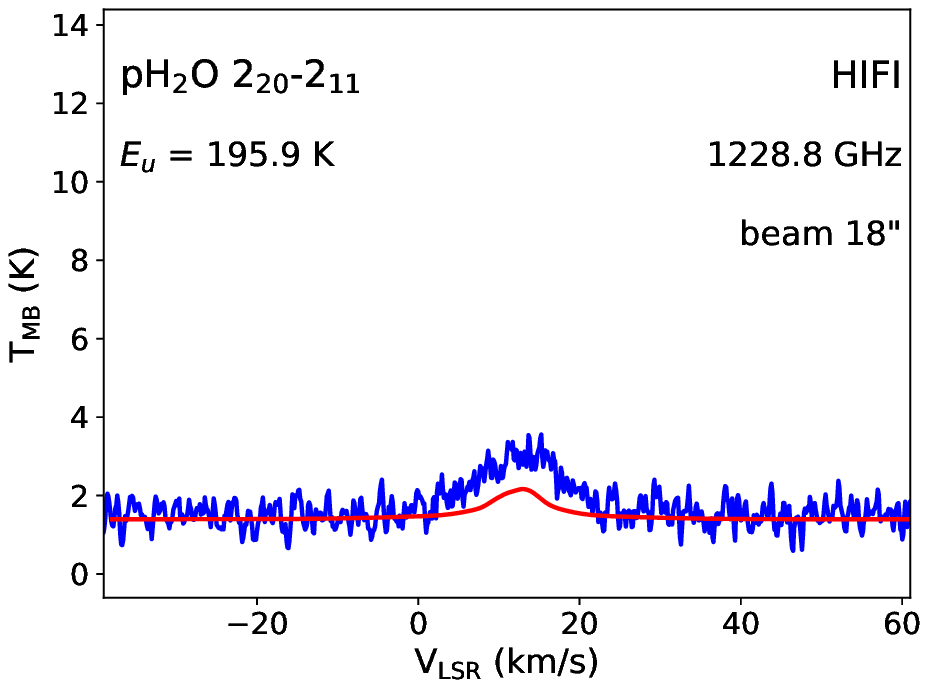}
\end{picture}}  

\end{picture}
\caption{Four lowest rotational lines of para-$\water$ observed towards OMC-2 FIR4 by {\sl Herschel}/HIFI in the course of the CHESS survey \citep{2010A&A...521L..22C}. Spectra simulated from the spherical shell model discussed in Sect.~\ref{sect_model} are shown in red.}
\label{ph2o_spectra}
\end{figure*}

%o-H2O
\begin{figure*}[htb]
\unitlength=1mm
\begin{picture}(160,110)(0,0)

\put(5,55){
\begin{picture}(0,0) 
  \includegraphics[width=7.3cm,angle=0]{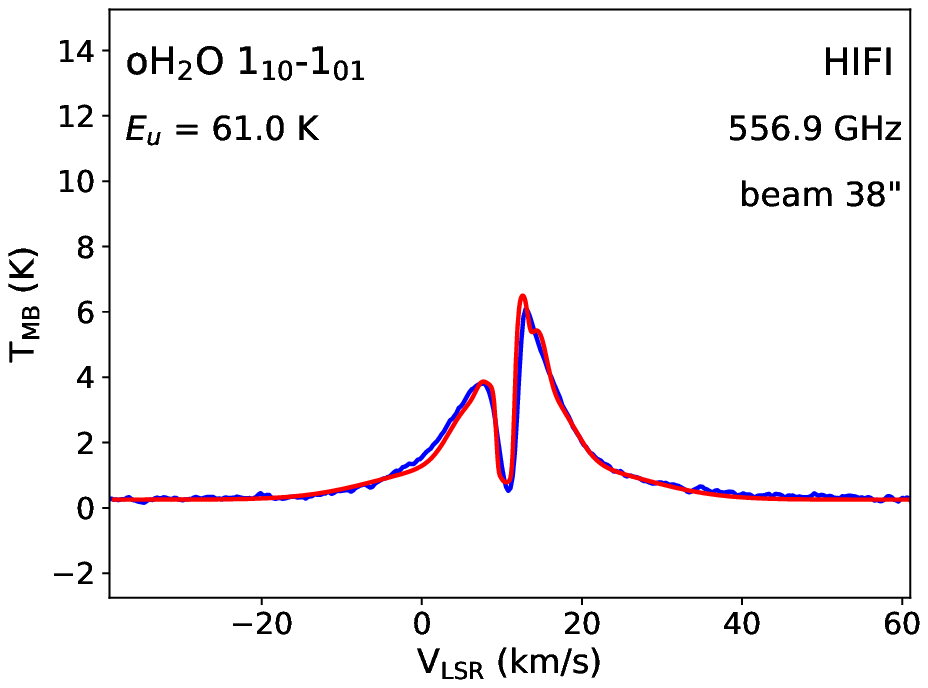}
\end{picture}}  

\put(90,55){
\begin{picture}(0,0) 
  \includegraphics[width=7.3cm,angle=0]{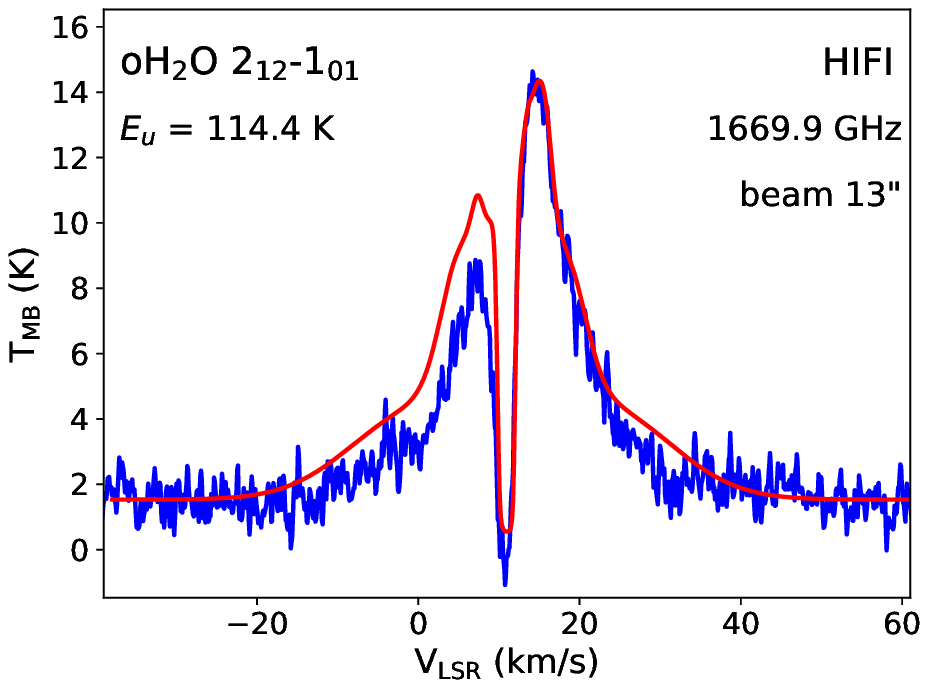}
\end{picture}}  

\put(5,0){
\begin{picture}(0,0) 
  \includegraphics[width=7.3cm,angle=0]{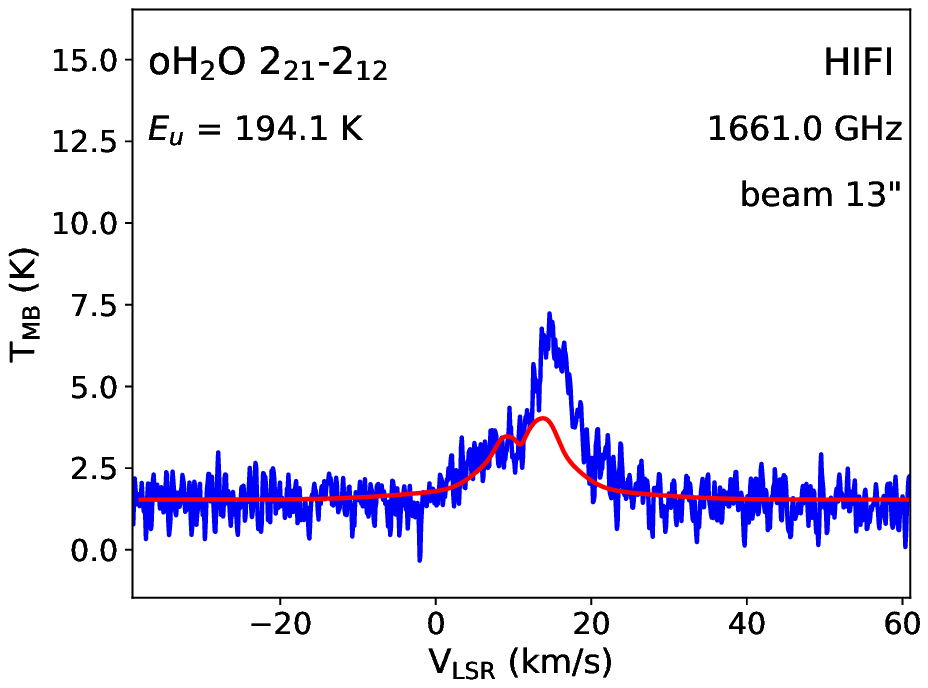}
\end{picture}}  

\put(90,0){
\begin{picture}(0,0) 
  \includegraphics[width=7.3cm,angle=0]{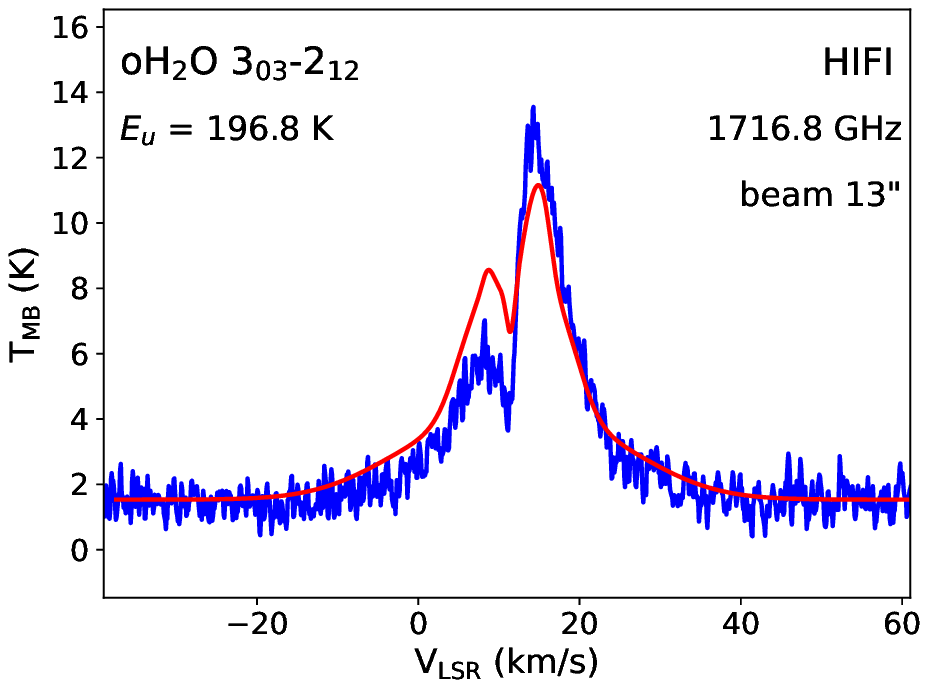}
\end{picture}}  

\end{picture}
\caption{Four lowest rotational lines of ortho-$\water$ observed towards OMC-2 FIR4 in the CHESS survey. The simulated spectra are shown in red.}
\label{oh2o_spectra}
\end{figure*}

%CO
\begin{figure*}[htb]
\unitlength=1mm
\begin{picture}(160,120)(0,0)

\put(0,60){
\begin{picture}(0,0) 
  \includegraphics[width=8cm,angle=0]{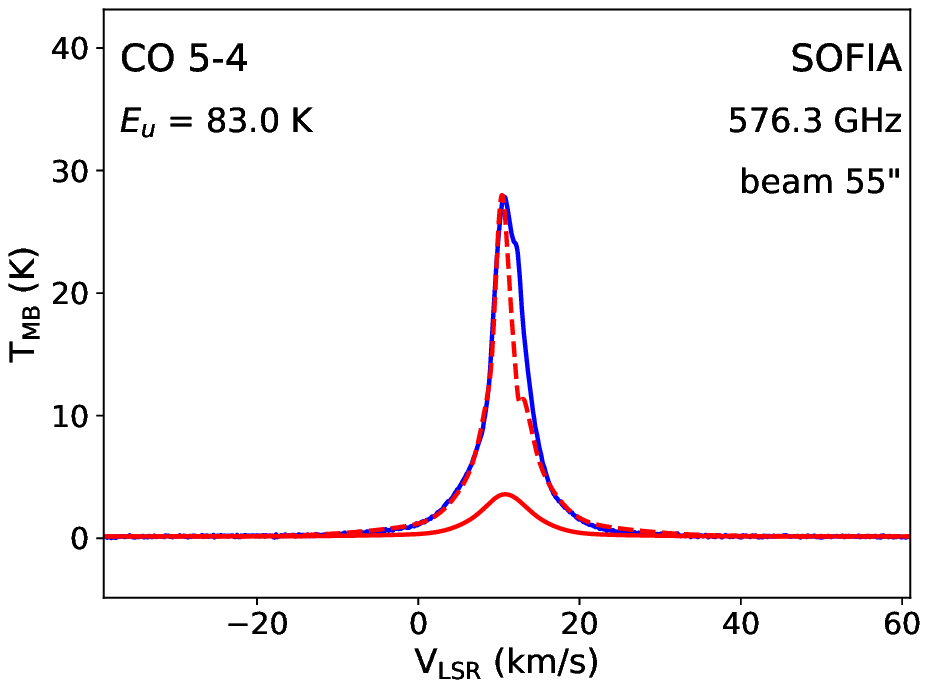}
\end{picture}}  

\put(85,60){
\begin{picture}(0,0) 
  \includegraphics[width=8cm,angle=0]{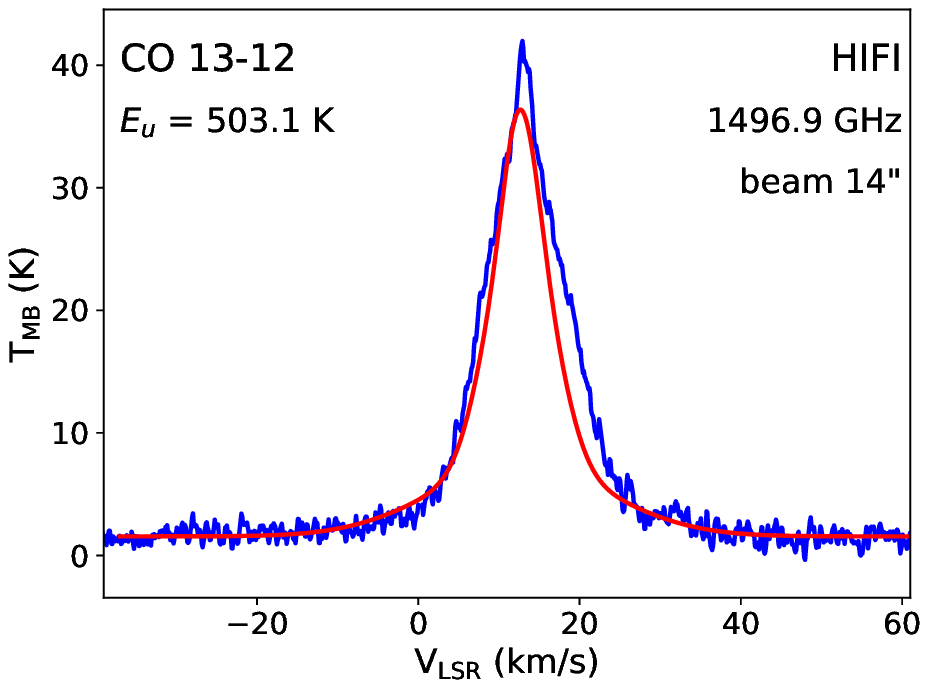}
\end{picture}}

\put(0,0){
\begin{picture}(0,0) 
\includegraphics[width=8cm,angle=0]{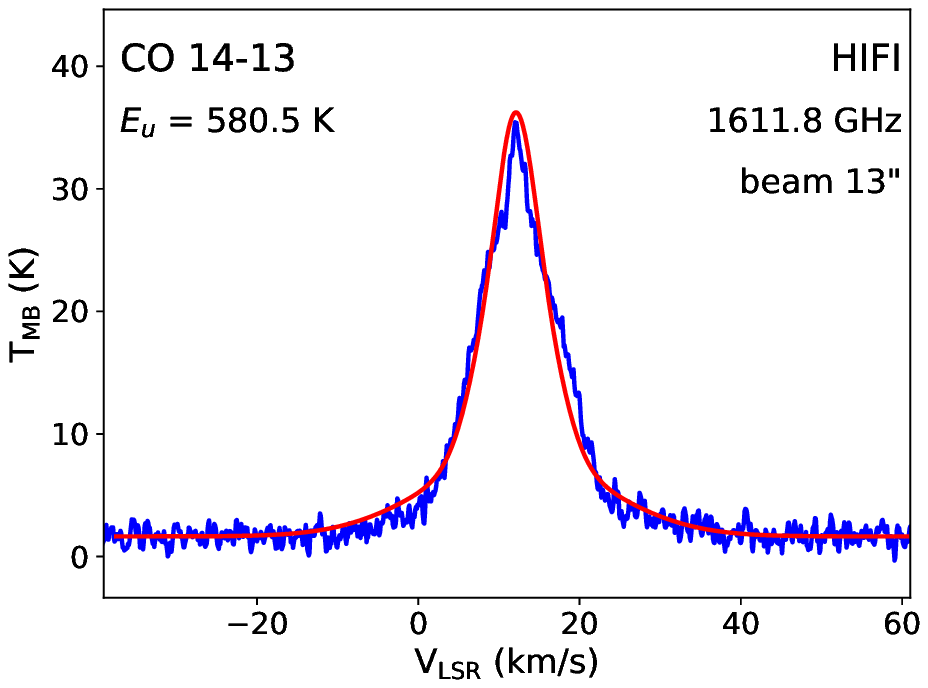}
\end{picture}}

\put(85,0){
\begin{picture}(0,0) 
\includegraphics[width=8cm,angle=0]{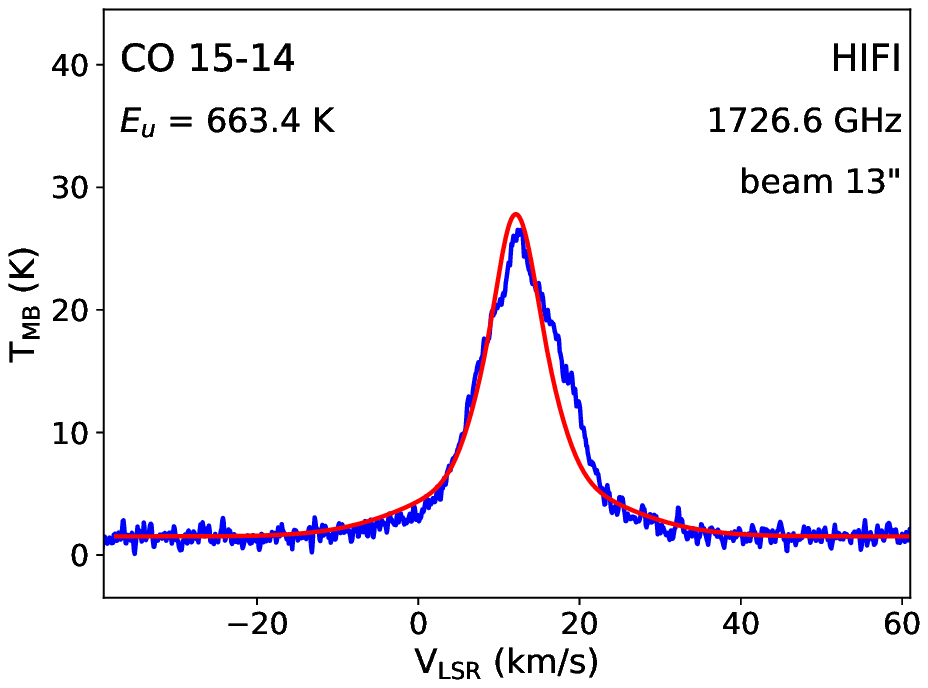}
\end{picture}}

\end{picture}
\caption{{\bf Top left:} The CO $J=5-4$ line observed with SOFIA. The spectrum calculated using the physical model and CO abundances given in Table~\ref{zones} is shown in red (solid line). The dashed red curve shows the simulated spectrum with the same model but with modified abundances in zones 1-3 and 5: $X({\rm CO})=5\times10^{-6}$, $10^{-5}$, $10^{-5}$, and $1.5\times10^{-5}$, respectively. {\bf Other panels:} Three {high-$J$} rotational lines of CO observed with Herschel/HIFI. Simulated spectra using the model and abundances in Table~\ref{zones} are shown in red.}
\label{co_spectra}
\end{figure*}

\section{Composition of model spectra}

The spherical shell model of the OMC-2 FIR4 clump consists of five zones described in Table~\ref{zones}. The contributions of these zones to the simulated \ion{O}{i}, OH, and HDO spectra are illustrated in Fig.~\ref{decomposition}. This figure shows spectra (coloured dashed curves) that would be observed if the abundance of the species were non-zero and equal to the value listed in Table~\ref{zones} in only one of the zones. When the emission and absorption of the line in other zones are taken into account,  the combined spectrum (black curve) is generally not equal to the sum of the spectra of the individual zones. 
\begin{figure*}[htb]
\unitlength=1mm
\begin{picture}(160,60)(0,0)

\put(120,0){
\begin{picture}(0,0) 
  \includegraphics[width=6cm,angle=0]{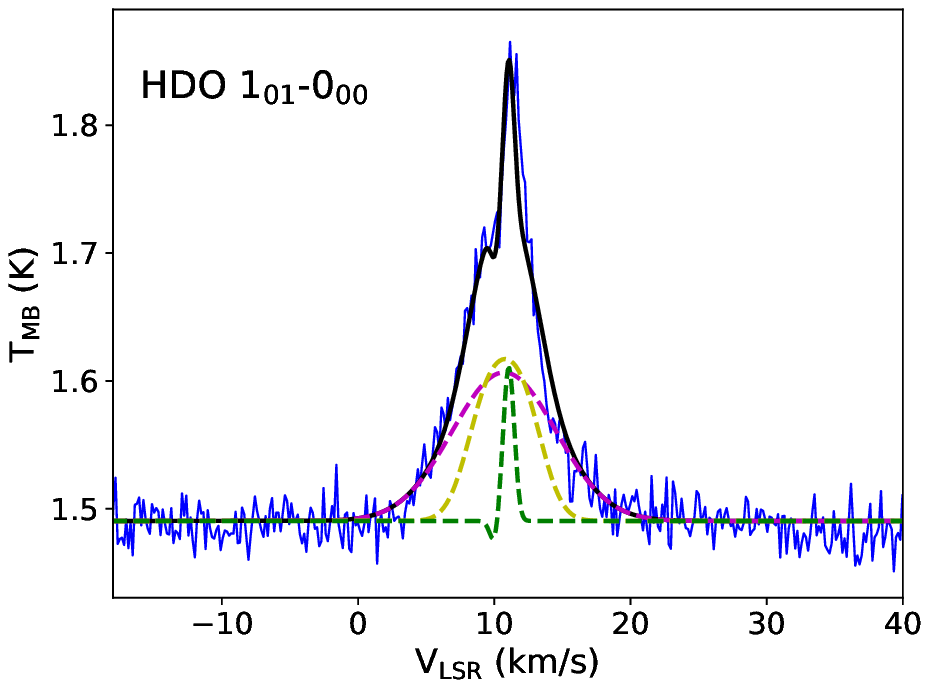}
\end{picture}}  
\put(60,0){
\begin{picture}(0,0) 
  \includegraphics[width=6cm,angle=0]{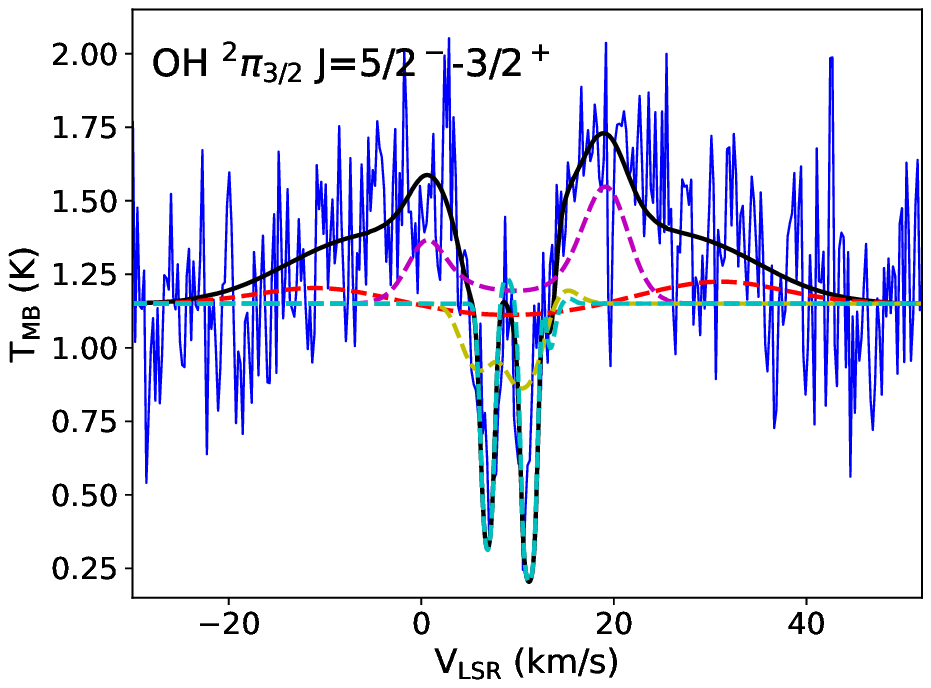}
\end{picture}}  
\put(0,0){
\begin{picture}(0,0) 
  \includegraphics[width=6cm,angle=0]{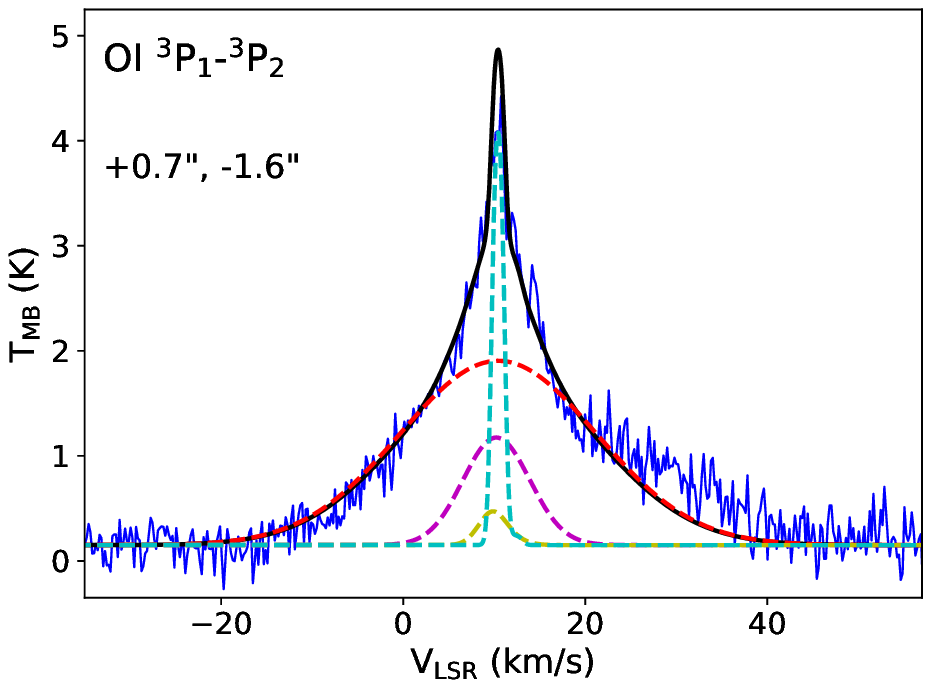}
\end{picture}}

\end{picture}
\caption{Contributions of different zones of the spherical shell model to the simulated \ion{O}{i}, OH, and HDO spectra. The following colour codes are used: red - zone 1, magenta - zone 2, yellow - zone 3, green - zone 4, and cyan - zone 5. The coloured spectra are calculated assuming that the abundance of the species is non-zero only in the zone in question, while the other zones do not emit or absorb in this line. Therefore, the emerging signal (black curve) can differ from the sum of the coloured spectra.}
\label{decomposition}
\end{figure*}

\section{Comparison between the [{C}{II}] and [{O}{I}] spectra}

{The [\ion{C}{ii}]\,158\,$\mu$m spectra towards the positions of our \ion{O}{i} observations are shown in Fig.~\ref{cii_oi_spectra}, together with the [\ion{O}{i}]\,63.2\,$\mu$m spectra. The \ion{C}{ii} data come from the large-scale survey of Orion A presented in \cite{2021A&A...652A..77H}, and are available at the Strasbourg astronomical Data Center (CDS)\footnote{http://cdsarc.u-strasbg.fr/viz-bin/cat/J/A+A/652/A77}. The narrow [\ion{O}{i}] line has LSR velocity and width similar to [\ion{C}{ii}] and its integrated intensity is approximately twice that of [\ion{C}{ii}]. This indicates that the narrow [\ion{O}{i}] line component originates in the PDR.} 

%CII & OI
\begin{figure*}[htb]
\unitlength=1mm
\begin{picture}(160,135)(0,0)
\put(33,0){
\begin{picture}(0,0) 
\includegraphics[width=5.5cm,angle=0]{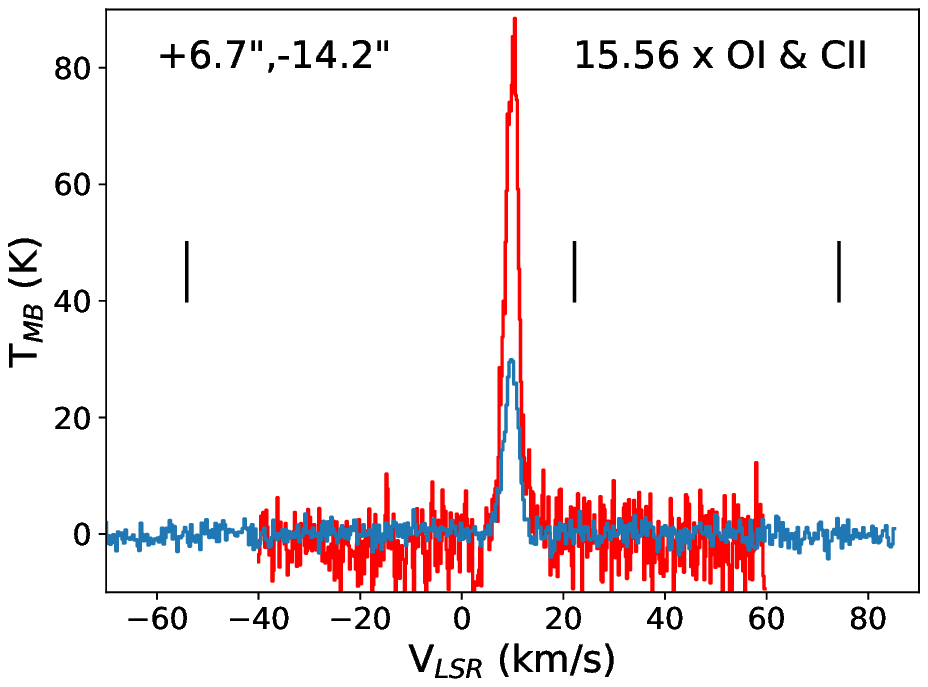}
\end{picture}}

\put(95,0){
\begin{picture}(0,0) 
\includegraphics[width=5.5cm,angle=0]{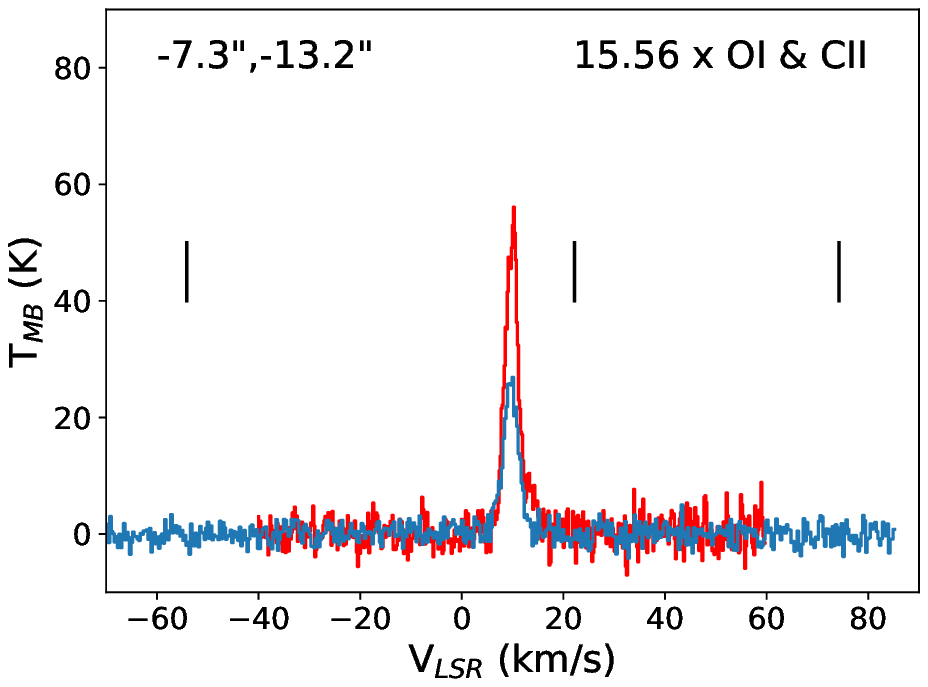}
\end{picture}}

\put(-5,50){
\begin{picture}(0,0) 
\includegraphics[width=5.5cm,angle=0]{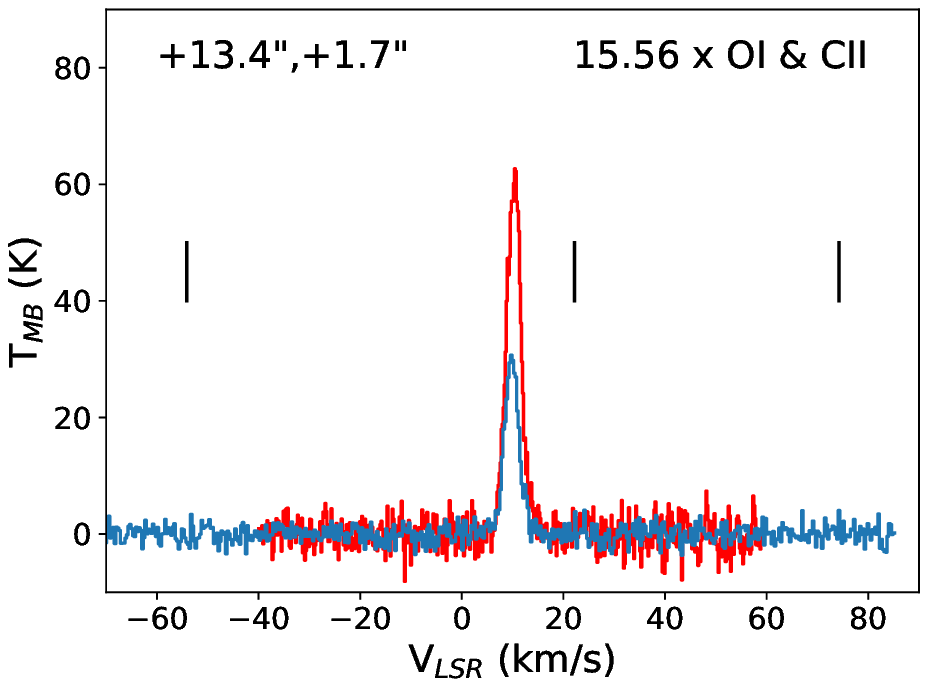}
\end{picture}}

\put(50,43){
\begin{picture}(0,0) 
\includegraphics[width=7cm,angle=0]{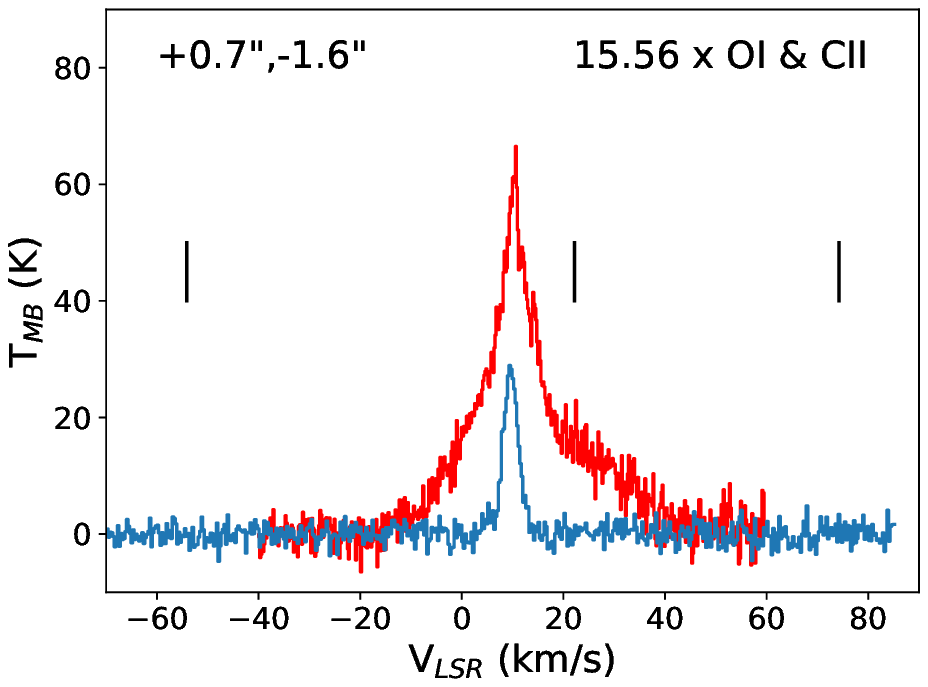}
\end{picture}}

\put(120,50){
\begin{picture}(0,0) 
\includegraphics[width=5.5cm,angle=0]{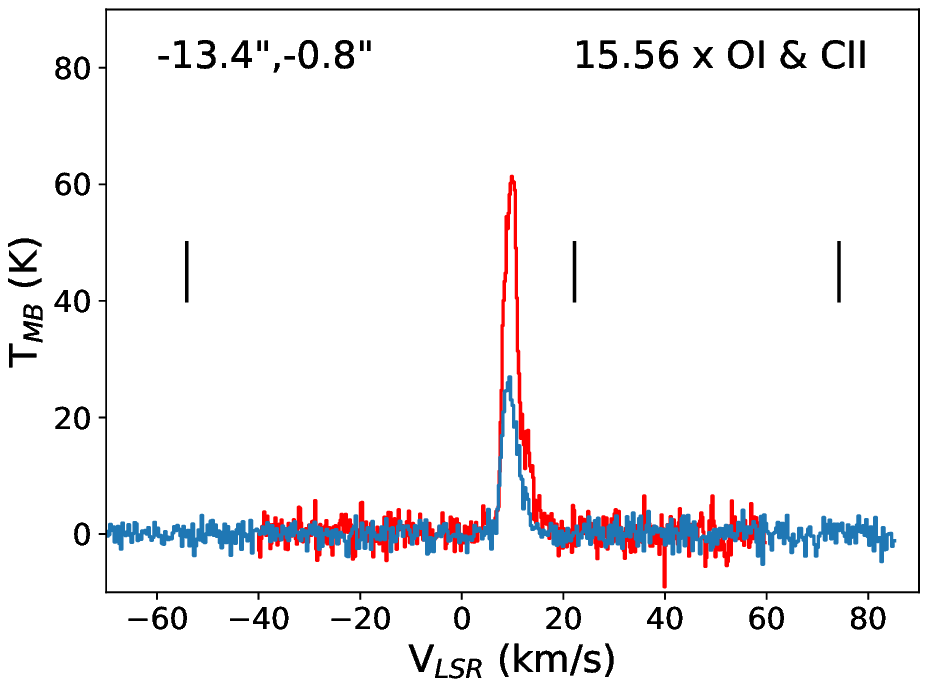}
\end{picture}}

\put(33,95){
\begin{picture}(0,0) 
\includegraphics[width=5.5cm,angle=0]{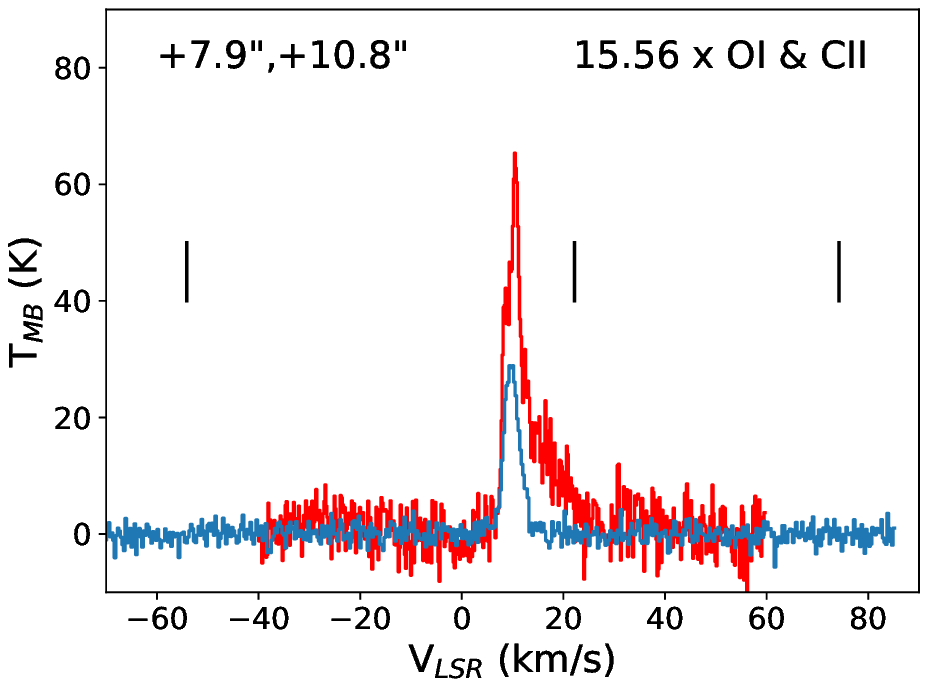}
\end{picture}}

\put(95,95){
\begin{picture}(0,0) 
\includegraphics[width=5.5cm,angle=0]{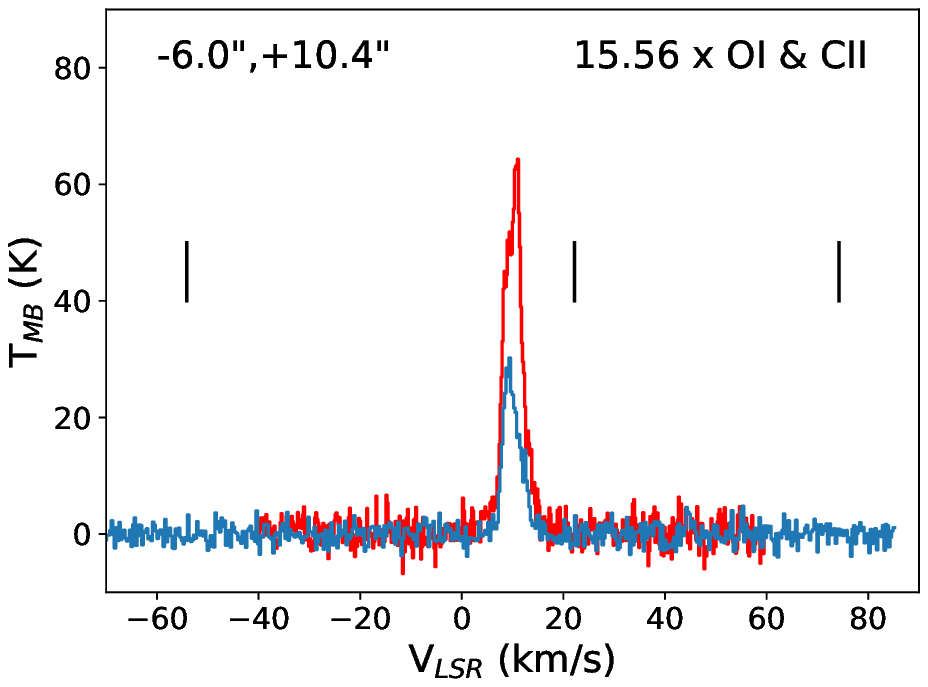}
\end{picture}}
\end{picture}
\caption{[{C}{II}]\,158\,$\mu$m (blue) and [{O}{I}]\,63.2\,$\mu$m (red) spectra towards OMC-2 FIR4 observed with SOFIA. The positions correspond to the pixels of the upGREAT HFA array. The [{O}{I}] spectra are multiplied by the factor $(\lambda_{\rm CII}/\lambda_{\rm OI})^3$, which makes the integrated intensities comparable when expressed in $\rm W\,m^{-2}\,sr^{-1}$. The black bars show the positions of the three hyperfine components of the  $^2P_{3/2}-^2P_{1/2}$ line of $\rm ^{13}{C}{II}$. These are not detected in the OMC-2 region.}
\label{cii_oi_spectra}
\end{figure*}

\end{appendix}
\end{document}